\documentclass[
    floats,
    floatfix,
    amssymb,
    prd,
    twocolumn,
    superscriptaddress,
    nofootinbib,
    unicode,
    colorlinks=true,
    linktocpage=true
]{revtex4-2}
\PassOptionsToPackage{dvipsnames,table}{xcolor}

\usepackage{graphicx, amssymb, amsmath, amsfonts, epsfig, fixmath}
\usepackage{epstopdf}
\usepackage[normalem]{ulem}
\usepackage{tabularx}
\usepackage{bm}
\usepackage{dcolumn}
\usepackage{rotating}
\usepackage{longtable}
\usepackage{enumerate}
\usepackage[unicode]{hyperref}
\usepackage{orcidlink}
\usepackage{slashed}
\usepackage{tensor, multirow}
\usepackage{url}
\usepackage{xcolor}
\usepackage[caption=false]{subfig}
\usepackage{float}
\usepackage{lipsum}

\hypersetup{
    colorlinks=true, 
    citecolor=MidnightBlue,
    linkcolor=MidnightBlue, 
    urlcolor=MidnightBlue, 
    linktocpage=true
}
\newcommand{\udot}{\,\underline{\cdot}\,}

\newcommand{\sissa}{SISSA, Via Bonomea 265, 34136 Trieste, Italy \&  INFN Sezione di Trieste}

\newcommand{\ifpu}{IFPU - Institute for Fundamental Physics of the Universe, Via Beirut 2, 34014 Trieste, Italy}

\newcommand{\centra}{CENTRA, Departamento de Física, Instituto Superior Técnico – IST,
Universidade de Lisboa – UL, Avenida Rovisco Pais 1, 1049-001 Lisboa, Portugal}

\newcommand{\imperial}{Department of Physics, Imperial College London, SW7 2AZ, London, United Kingdom}

\newcommand{\bicocca}{Dipartimento di Fisica “G. Occhialini", Università degli Studi di Milano-Bicocca, Piazza della Scienza 3, I-20126 Milano, Italy}

\newcommand{\infnmilano}{INFN, Sezione di Milano-Bicocca, Piazza della Scienza 3, I-20126 Milano, Italy}

\begin{document}

\title{Searching for Gravitational Waves with Gaia and its Cross‑Correlation with PTA: Absolute vs Relative Astrometry}

\author{Massimo Vaglio~\orcidlink{0000-0002-7285-3489}}
\email{mvaglio@sissa.it}
\affiliation{\sissa}
\affiliation{\ifpu}

\author{Mikel Falxa~\orcidlink{0000-0002-0815-1781}}
\email{mikel.falxa@unimib.it}
\affiliation{\bicocca}
\affiliation{\infnmilano}

\author{Giorgio Mentasti~\orcidlink{0000-0003-1115-9220}}
\email{g.mentasti21@imperial.ac.uk}
\affiliation{\imperial}

\author{Arianna I. Renzini~\orcidlink{0000-0002-4589-3987}}
\email{arianna.renzini@unimib.it}
\affiliation{\bicocca}
\affiliation{\infnmilano}

\author{Adrien Kuntz~\orcidlink{0000-0002-4803-2998}}
\email{adrien.kuntz@tecnico.ulisboa.pt}
\affiliation{\centra}

\author{Enrico Barausse~\orcidlink{0000-0001-6499-6263}}
\email{barausse@sissa.it}
\affiliation{\sissa}
\affiliation{\ifpu}

\author{Carlo R. Contaldi~\orcidlink{0000-0001-7285-0707}}
\email{c.contaldi@imperial.ac.uk}
\affiliation{\imperial}

\author{Alberto Sesana~\orcidlink{0000-0003-4961-1606}}
\email{alberto.sesana@unimib.it}
\affiliation{\bicocca}
\affiliation{\infnmilano}

% Abstract
%%%%%%%%%%%%%%%%%%%%%%%%%%%%%%%%%%%%%
\begin{abstract}
Astrometric missions like Gaia provide exceptionally precise measurements of stellar positions and proper motions. Gravitational waves traveling between the observer and distant stars can induce small, correlated shifts in these apparent positions, a phenomenon known as astrometric deflection. The precision and scale of astrometric datasets make them well-suited for searching for a stochastic gravitational wave background, whose signature appears in the two-point correlation function of the deflection field across the sky.
Although Gaia achieves high accuracy in measuring angular separations in its focal plane, systematic uncertainties in the satellite’s absolute orientation limit the precision of absolute position measurements. These orientation errors can be mitigated by focusing on relative angles between star pairs, which effectively cancel out common-mode orientation noise.
In this work, we compute the astrometric response and the overlap reduction functions for this relative astrometry approach, correcting previous expressions presented in the literature. We use a Fisher matrix analysis to compare the sensitivity of relative astrometry to that of conventional absolute astrometry. Our analysis shows that while the relative method is theoretically sound, its sensitivity is limited for closely spaced star pairs within a single Gaia field of view. Pairs with large angular separations could provide competitive sensitivity, but are practically inaccessible due to Gaia's scanning law.
Finally, we demonstrate that combining astrometric data with observations from pulsar-timing arrays leads to slight improvements in sensitivity at frequencies $\gtrsim 10^{-7}$ Hz.

\end{abstract}
%%%%%%%%%%%%%%%%%%%%%%%%%%%%%%%%%%%%%%%

\maketitle

\section{Introduction}
Gravitational-wave (GW) astronomy is significantly advancing our understanding of astrophysics and cosmology. Current ground-based detectors—LIGO, Virgo, and KAGRA—routinely observe binary mergers involving astrophysical compact objects, identified as black holes (BHs) or neutron stars (NSs)~\cite{KAGRA:2021vkt}. These observations provide valuable insights into their masses and spins, as well as their merger rates and distances, which are essential for understanding stellar evolution and the formation of cosmic structures.

Beyond individual and transient events, a key focus of GW searches is the stochastic gravitational wave background (SGWB). If present, this background would arise from the superposition of numerous unresolved and uncorrelated sources, which could have either astrophysical~\cite{Regimbau:2011rp} or cosmological~\cite{Caprini:2018mtu} origins. Astrophysical contributions to the background include GWs from compact binaries~\cite{Jaffe:2002rt, Sesana:2008mz}, core-collapse supernovae~\cite{Ferrari:1998ut,Finkel:2021zgf}, and rotational instabilities of NSs~\cite{Surace:2015ppq}, typically producing a power-law spectrum. In contrast, a cosmological background may originate from processes in the early universe, e.g. inflation~\cite{Starobinsky:1979ty,Allen:1987bk}, reheating~\cite{Easther:2006gt}, cosmic strings~\cite{Vilenkin:2000jqa}, or the QCD phase transition~\cite{Caprini:2015zlo}.

Pulsar timing array (PTA) collaborations, such as the North American Nanohertz Observatory for Gravitational Waves (NANOGrav) \cite{NANOGRAV}, the European Pulsar Timing Array (EPTA) \cite{EPTA}, the MeerKAT Pulsar Timing Array (MPTA) \cite{mpta_data}, and the International Pulsar Timing Array (IPTA) \cite{IPTA}, are currently probing extremely low frequencies in the nHz range. Future space-based missions, such as the Laser Interferometer Space Antenna (LISA)~\cite{LISA:2017pwj,LISA:2024hlh}, will instead focus on the SGWB in the mHz band. In June 2023, evidence was reported of a common-spectrum process with a strain amplitude of approximately 
$h \sim 2\text{--}3\times 10^{-15}$, observed with a $2\text{--}4\sigma$ significance depending on the dataset, 
just below the $5\sigma$ detection threshold~\cite{NANOGrav:2023gor, EPTA:2023fyk, Reardon_2023, Xu_2023, mpta_gwb}. This process is consistent with an SGWB originating from the coalescence of supermassive black hole binaries (SMBHB), which can form during galaxy mergers and are expected to dominate the signal in the PTA band~\cite{david_confronting_mbh_nhz, NANOGrav:2023hfp, Phinney:2001di, epta_iv_implications}. However, further investigations are essential to confirm this tentative detection and to explore its astrophysical and cosmological implications.

The detection of the SGWB with PTAs leverages the correlated timing residuals in the arrival times of radio pulses from millisecond pulsars, which are influenced by GWs. When a signal is present, this correlation exhibits a specific dependence on the angular separation between pulsars, described by the Hellings-Downs curve~\cite{Hellings:1983fr}. The use of PTAs and this correlation pattern to constrain a GW background was already demonstrated in the 1980s~\cite{Backer:1983, Hellings:1983fr,Davis:1985, 1990ApJ...361..300F, Rajagopal:1994zj}. %\mv{I think these are the first works that refer to PTA, but maybe do a double check.} 
By contrast, the idea of probing the SGWB through astrometric measurements has been considered more recently, with a detailed investigation first presented in Ref.~\cite{PhysRevD.83.024024}, and further explorations in the context of precision astrometry missions like Gaia, later on~\cite{Moore:2017ity, Klioner:2017asb, OBeirne:2018slh, Darling:2018hmc, Jaraba:2023djs, Caldarola:2024uig}, including synergy with PTA~\cite{Caliskan:2023cqm, Cruz:2024diu}. This approach relies on the fact that GWs also induce, alongside a frequency shift, an astrometric deflection of electromagnetic signals ---a change in the apparent angular positions of stars relative that is transverse to the unperturbed line-of-sight~\cite{Braginsky:1990ps, Pyne:1995iy, Kaiser_1997, Kopeikin:1998ts}. This effect also carries a characteristic angular cross-correlation pattern. 

In Gaia, the positions of stars are measured relative to the satellite’s orientation (attitude). However, determining their absolute positions requires comparing observations taken at different times, carefully accounting for the satellite’s motion. Because the satellite’s orientation at any given time is known with less precision than the stars’ relative positions, the analysis of Gaia data involves a joint fit that simultaneously solves for all the source parameters, the satellite attitude, and various calibration factors~\cite{2012A&A...538A..78L, lindegren2021gaia}. As a result, the absolute astrometric measurements inherently carry systematic uncertainties from this complex procedure. These uncertainties can be mitigated by shifting from absolute to {\sl relative} astrometry, a promising technique that has been employed in electromagnetic observations~\cite{zverev1976importance}, and more recently proposed for GW detection using photometric surveys~\cite{Wang:2020pmf, Wang:2022sxn,Pardo:2023cag, Zhang:2025srs} as well as Gaia astrometric data~\cite{Crosta:2024udx}. By examining pairs of stars and focusing on changes in their relative angular separations, rather than in their absolute position, one can bypass many of the systematics associated with the simultaneous satellite attitude reconstruction, achieving more precise relative measurements. 

In this work, we present a comparison between absolute and relative approaches. To achieve this, we derive the GW response function for relative astrometry and the associated overlap-reduction function (ORF), correcting previous expressions presented in the literature~\cite{Crosta:2024udx}. Then we perform a Fisher matrix forecast to evaluate astrometric performance across different setups, comparing and combining them with PTAs. Similar analyses combining PTA and astrometric datasets were performed in ~\cite{Caliskan:2023cqm, Cruz:2024diu} using different noise models and dataset properties.

The paper is structured as follows. In Section~\ref{sec:astrometric_response_tot}, we review the astrometric effect and introduce relative astrometry. In Section~\ref{sec:response_to_sgwb_tot}, we characterize the response to a SGWB and derive the ORF for relative astrometry and relative astrometry-pulsar correlations. In Section~\ref{sec:Fisher_forecast}, we present the theoretical framework underlying our analysis and describe the Fisher matrix formalism used to compare absolute and relative astrometry. The main results are discussed in Section~\ref{sec:results}, and we summarize our main conclusion and outlook in Section~\ref{sec:conclusions}. The appendices provide additional details about the theoretical framework. We introduce a simple derivation of the astrometric shift in Appendix~\ref{app:astrometric_shift}, which highlights its complementarity with the GW-induced pulsar time delay. In Appendix~\ref{app:differential_shift}, we directly compare our expression for the relative astrometric response with that of Ref.~\cite{Crosta:2024udx}, showing that the discrepancy arises from a missing contribution in their result, whose origin and justification we explicitly trace and clarify. Finally, in Appendix~\ref{app:transmission_function} we show an analytical derivation of the transmission function formalism used in Sec.~\ref{sec:Fisher_forecast} and originally introduced in \cite{Hazboun_2019}. This formalism is designed to filter out the effect of first order errors in the timing (or astrometric) model fit at very low frequencies and is essential to perform a reasonable forecast.

\section{Astrometric response to a gravitational wave}
\label{sec:astrometric_response_tot}
Astrometric searches for GWs are based on the fact that gravitational perturbations can alter the light path from distant stars, leading to measurable shifts in their apparent positions in the sky. In the following, we discuss the astrometric response for a single star and for a pair of stars. 
\subsection{Absolute astrometry: response of a single star}
\label{sec:astrometric_response}
The astrometric deflection effect can be quantitatively described as follows. A photon propagating from a distant source to Earth travels in a spacetime $g_{\mu\nu}=\eta_{\mu\nu}+h_{\mu\nu}+\mathcal{O}(h^2)$, where $\eta_{\mu\nu}$ is the flat background metric and $h_{\mu\nu}$ represents a gravitational perturbation. We can decompose the photon wave vector $\sigma$ onto a tetrad basis $e^{(\mu)}_\alpha$, associated with the observer's locally inertial reference frame, for the perturbed metric
\begin{equation}
\sigma^\alpha=\nu e_{(0)}^\alpha +\nu n^i e_{(i)}^\alpha\,,
\label{eq:photon_wave_vector2}
\end{equation} 
where $e^{(\mu)}_{(0)}\equiv u^{\mu}$ is the observer four-velocity, $\{e^{(\mu)}_{(i)}\}_{i=1,2,3}$ are the spatial directions in the observer's frame and $g_{\mu\nu}e^{(\mu)}_\alpha e^{(\nu)}_\beta=\eta_{\alpha \beta}$. The change in the star's position results in a small periodic shift in the normalized spatial components $n^i$ of the photon wave vector, which define the direction of the incoming light ray with respect to the observer.
If $h_{\mu\nu}$ is a plane GW with frequency $\omega$ propagating in the direction $\bf p$, then in the limit where a star lies many wavelength away from the Earth, i.e. when $\omega \tau\gg 1$, where $\tau$ is the light time of travel, the astrometric deflection can be expressed as~\cite{PhysRevD.83.024024,Pyne:1995iy}
\begin{equation}
\delta n^i = \left(\frac{n^i  +p^i}{2 (1 + \mathbf{n} \cdot \mathbf{p})} n^j n^k - \frac{1}{2} \delta^{ik}n^j\right)  h^E_{jk}\,,
\label{eq:astrometric_shift}
\end{equation}
where $h^E_{ij}$ corresponds to the GW perturbation tensor evaluated at the Earth location. In Appendix~\ref{app:astrometric_shift}, we present a simple derivation of Eq.~\eqref{eq:astrometric_shift} that only makes use of conservation equations, avoiding the direct integration of the geodesics adopted in previous approaches~\cite{Pyne:1995iy, PhysRevD.83.024024}.
 
 Eq.~\eqref{eq:astrometric_shift} is the analogue of the GW-induced frequency shift $\delta \nu$ (redshift) in pulsar timing, which leads to the observed variations in pulse arrival times. Note, however, that the shift $\delta \mathbf{n}$ is a vector and has two degrees of freedom, as the normalization forces the variation to be orthogonal to the propagation direction. Because $n^i$ are spatial components in the tetrad frame, the scalar product is with respect to the flat metric $\delta_{ij}$, i.e. $\mathbf{n} \cdot \mathbf{n}$ and $\delta \mathbf{n} \cdot \mathbf{n}=0$. The two independent components of $\delta \mathbf{n}$ can be mapped in a shift in the two angles that locate a star in the sky, $\delta \theta$ and $\delta \phi$, where $\theta$ and $\phi$ identify the star unperturbed sky position. 
 
 The expression in Eq.~\eqref{eq:astrometric_shift} represents the change in the apparent sky position of an individual star whose unperturbed direction is $\mathbf{n}$, and we will refer to an astrometric search based on it as \emph{absolute astrometry}. A shortcoming of this approach is that any practical use of Eq.~\eqref{eq:astrometric_shift} assumes that the reference frame (tetrad) is known for all the observed stars. Astrometric survey from satellites like Gaia will, in general, be composed of many observations taken with the satellite in different orientations. In particular, Gaia is composed of two telescopes, each having a $\sim 0.7$ deg field of view (FoV) and separated by an angle (basic angle) of $106.5$ deg~\cite{Gaia:2016zol}. The FoVs rotate around an orthogonal axis that is precessing around the direction pointing at the Sun, while the whole satellite moves along its orbit~\footnote{See, for instance, Fig.~6 in~\cite{Gaia:2016zol} and the discussion in~\cite{2012A&A...538A..78L}.}. The motion ensures that the entire sky is mapped within a period of $\sim 6$ months. An \emph{astrometric solution} is then fitted from the data, accounting for the change in the satellite attitude over time~\cite{2012A&A...538A..78L}. This procedure introduces correlations and uncertainty in the measured sky position, which are expected to be larger compared to the precision achieved by a single observation in one FoV~\cite{Holl:2012A&A543A15,geyer2025influencecontinuousplanegravitational}. 
 
\subsection{Relative astrometry: response of a pair of stars}

 Relative astrometry avoids the need to fit for the satellite attitude by considering as the fundamental observable the relative angle between pairs of stars and their astrometric response to a GW. This approach was recently considered for GW searches with Gaia in Ref.~\cite{Crosta:2024udx}. Given two stars with directions $\mathbf{n}_1$ and $\mathbf{n}_2$, the cosine of their relative angle is given by the scalar product $\cos{\psi_{12}}=\mathbf{n}_1 \cdot \mathbf{n}_2$. The $\mathcal{O}(h)$ variation, due to the GW is
\begin{equation}
\delta \cos{\psi_{12}} = \delta \mathbf{n}_1 \cdot \mathbf{n}_{2} + \mathbf{n}_{1} \cdot \delta \mathbf{n}_2\,,
\label{eq:delta_cos}
\end{equation}
where $\mathbf{n}_{I=1,2}$ are the unperturbed directions. By substituting Eq.~\eqref{eq:astrometric_shift} into Eq.~\eqref{eq:delta_cos} we obtain
\begin{align}
\delta\cos{\psi_{12}}%=&\delta\ell_0^1\ell_0^2+\ell_0^1 \delta \ell_0^2+h_{ij}\tilde{\ell_0}^{1i}\tilde{\ell_0}^{2j} 
=&\frac{\mathbf{n}_{1}\cdot \mathbf{n}_{2}  +\mathbf{p} \cdot \mathbf{n}_{2}}{2 (1 + \mathbf{n}_{1} \cdot \mathbf{p})} h_{jk} n_{1}^{j} n_{1}^{k}\nonumber \\ 
&+ 
\frac{\mathbf{n}_{1} \cdot \mathbf{n}_{2}  +\mathbf{p} \cdot \mathbf{n}_{1}}{2 (1 + \mathbf{n}_{2} \cdot  \mathbf{p})} h_{jk} n_{2}^{j} n_{2}^{k}\,, 
- h_{ij} n_{1}^{i} n_{2}^{j}
\label{eq:delta_cos_expl}
\end{align}
where, again, the $\mathbf{n}_I$ and their components are zero-order in $h$. Note that this expression (since for small angles $\psi_{12}$ one has $\delta\cos{\psi_{12}}\sim -\delta \psi_{12} \sin{\psi_{12}}$) amends Eq.~(12) of~\cite{Crosta:2024udx} for $\delta\cos{\psi_{12}}$, which lacks a term $- h_{ij} n_{1}^{i} n_{2}^{j}$ on the right-hand side. The oversight of Ref.~\cite{Crosta:2024udx} is discussed in more detail in Appendix~\ref{app:differential_shift}.

If one considers star pairs lying in the focal plane of a single Gaia FoV at a given time, then their angular separation will always be small. One can then expand in orders of the unperturbed angle $\psi_{12}$ taking $\mathbf{n_1}=\mathbf{n}$ and $\mathbf{n}_2\sim\mathbf{n}+\mathbf{n}_{12}+\mathcal{O}(\psi_{12}^2)$. The relative change in $\psi_{12}$ becomes in this limit
\begin{align}
&\delta \psi_{12}= \psi_{12}\,\frac{h_{ij}}{2} \left[v^iv^j-\frac{n^i n^j}{(1+\mathbf{n}\cdot \mathbf{p})}\right]+\mathcal{O}(\psi_{12}^3)\,, \nonumber \\ 
&v^i\equiv \hat{n}_{12}^i-\frac{(\hat{\mathbf{n}}_{12} \cdot \mathbf{p}) n^i}{(1+\mathbf{n}\cdot \mathbf{p})}\,,
\label{eq:small_angle}
\end{align}
where $\hat{\mathbf{n}}_{12}\equiv\mathbf{n}_{12}/|\mathbf{n}_{12}|$.
This result shows that $\delta \psi_{12}$ vanishes in the limit of zero angular separation, $\psi_{12} \sim 0$. Analogously to the response of linear interferometers, where $\delta_L \propto h \cdot L$, one finds that for small $\psi_{12}$ the astrometric effect behaves as $\delta \psi_{12} \propto h \cdot \psi_{12}$ (cf. Appendix~\ref{app:differential_shift}).
From Eq.~\eqref{eq:small_angle}, it is also evident that $\delta \psi_{12}$ depends not only on the unperturbed propagation direction $\mathbf{n}$, but also on the orientation $\hat{\mathbf{n}}_{12}$ of the star separation vector projected onto the plane tangent to the celestial sphere at the star location. 

\section{Response to a stochastic background}
\label{sec:response_to_sgwb_tot}
We now want to characterize the astrometric response to an SGWB. The latter can be described as a superposition of many plane waves with all possible polarizations and whose amplitude coefficients are random variables, specified by some statistical properties, i.e.
\begin{align}
h_{\mu\nu}(t,\mathbf{x})  = & \sum_{P=+,\times}  \int df \int d\Omega_{\mathbf{p}}
\, h_P(f,\mathbf{p}) e^{2i\pi f t} \nonumber
\\
&\qquad \qquad \times e^{-2i\pi f \mathbf{p}\cdot\mathbf{x}}
\, e^A_{\mu\nu}(\mathbf{p})
+ \text{c.c.}\,,
\end{align}
where $f$ and $\mathbf{p}$ are the frequency and direction of each GW mode and ${e^P_{\mu\nu}(\mathbf{p})}_{P=+,\times}$ are the polarization tensors. Assuming the background to be stationary, with zero mean, isotropic and unpolarized, the mode amplitudes $h_P(f,\mathbf{p})$ satisfy
\begin{equation}
\langle h_P(f,\mathbf{p})h^*_{P'}(f',\mathbf{p}')\rangle=S_h (f)\delta(f-f')\delta_{PP'} \delta^2(\mathbf{p},\mathbf{p}')\,,
\end{equation}
where the brackets $\langle . \rangle$ denote the ensemble average and $S_{h}(f)$ is the power spectral density of the SGWB. Because the response (cf. Eqs.~\eqref{eq:astrometric_shift}-\eqref{eq:delta_cos_expl}) is linear in the perturbation, the properties of the SGWB are fully described by the two-point correlation function of the astrometric deflection. 
\subsection{Absolute astrometry ORF}
\label{sec:response_to_sgwb}
\begin{figure}[thp]
\centering
\includegraphics[width=0.5\textwidth]{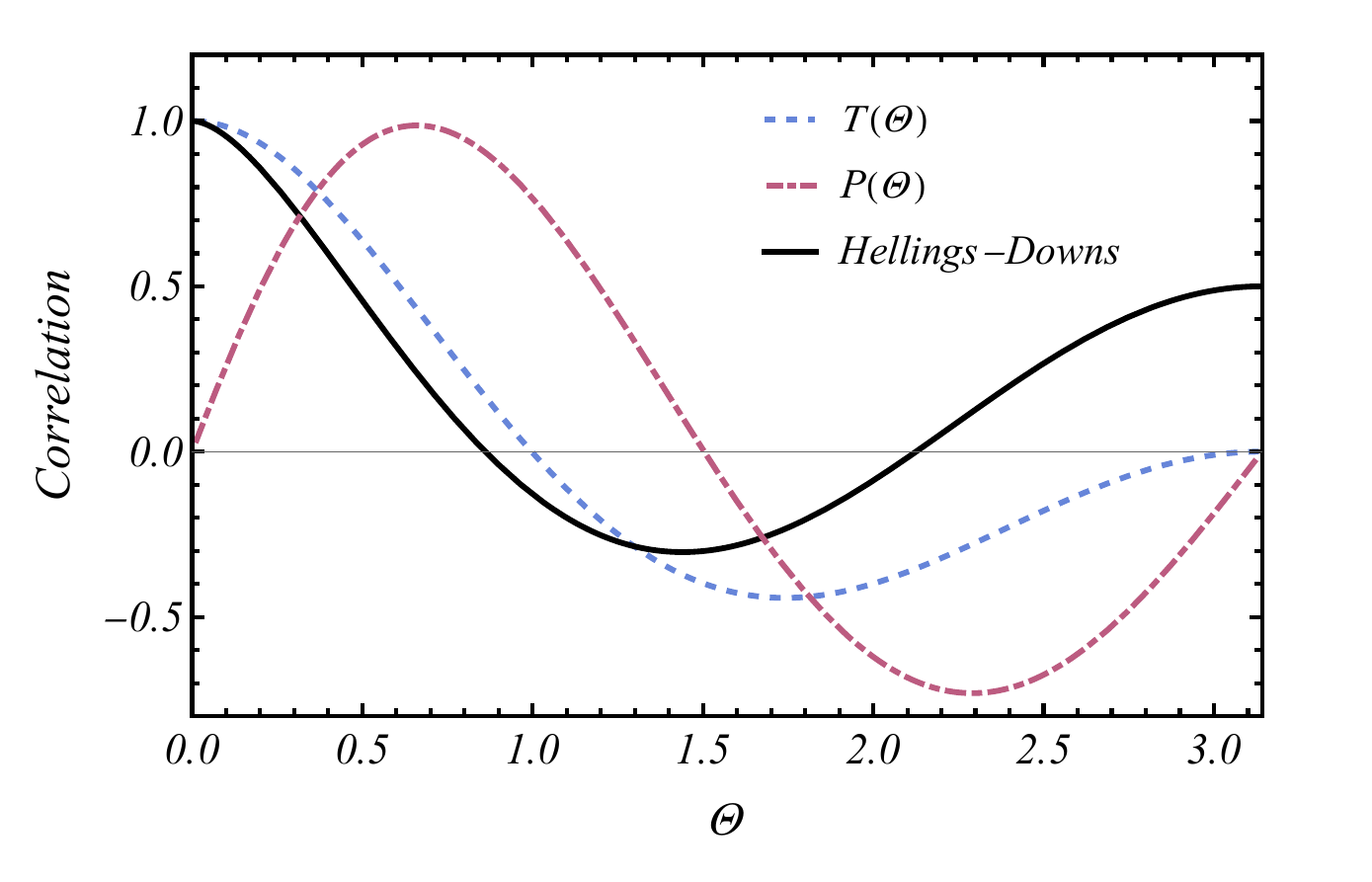}
    \caption{
Comparison of the astrometric ORF (blue, dashed), given by the single expression $\mathcal{T}(\Theta)$, the cross-correlation ORF (purple, dot-dashed)  $\mathcal{P}(\Theta)$ (also defined in Ref.~\cite{Mihaylov:2018uqm}), with the Hellings–Downs curve (black, solid), which characterizes redshift correlations. Curves are shown as functions of the angular separation on the sky, $\Theta$. The normalization has been chosen so that they each have a maximum value of one (neglecting the pulsar/star term).
}
    \label{fig:hd_Mihaylov}
\end{figure}

In the case of absolute astrometry, the expectation value of the product of astrometric responses takes the form 
\begin{equation}
\langle \delta n_a^i \delta n_b^j\rangle \propto \mathcal{H}^{ij}_A(\mathbf{n}_a,\mathbf{n}_b)\times S_h(f)\,,
\label{eq:Mihaylov}
\end{equation}
where $\mathcal{H}^{ij}_A(\mathbf{n}_a,\mathbf{n}_b)$ is the ORF, that depends only on the star's locations and $\mathbf{n}_a$ and $\mathbf{n}_b$ are the directions to two stars. By rewriting Eq.~\eqref{eq:astrometric_shift} as $\delta n^i = K^{ijk}(\mathbf{n},\mathbf{p})h^E_{jk}$ the ORF can be defined as
\begin{equation}
\mathcal{H}_A^{ij}(\mathbf{n}_a,\mathbf{n}_b)=\!\!\!\!\sum_{A=+,\times}\!\int d\Omega_\mathbf{p} K^{ikh}(\mathbf{n}_a,\mathbf{p})e^A_{kh} K^{jlm}(\mathbf{n}_b,\mathbf{p})e^A_{lm}\,,
\label{eq:orf_def}
\end{equation}
i.e. as the sky average of the product of the geometric part of the response functions. Expressions like Eq.~\eqref{eq:orf_def} have been computed by many authors~\cite{PhysRevD.83.024024, Gair:2015hra, OBeirne:2018slh, Mihaylov:2018uqm, Mentasti:2023gmr, Inomata:2024kzr,Allen:2024bnk}. In Ref.~\cite{Mihaylov:2018uqm} $\mathcal{H}^{ij}_A(\mathbf{n}_a,\mathbf{n}_b)$ is expressed as 
\begin{equation}
\mathcal{H}^{ij}_A= \mathcal{T}(\Theta)U^{ij}(\mathbf{n}_a,\mathbf{n}_b)\,,
\label{eq:Mihaylov_bis}
\end{equation}
where $\mathcal{T}(\Theta)$ is a scalar function, analogous to the Hellings-Downs, that only depends on the separation angle between the two stars $\Theta=\arccos{(\mathbf{n}_a\cdot \mathbf{n}_b)}$, while $U^{ij}(\mathbf{n}_a,\mathbf{n}_b)$ is a function of the unperturbed star directions that depends on the coordinate choice. Following~\cite{Mihaylov:2018uqm} we can construct $U^{ij}(\mathbf{u},\mathbf{v})$ for two unit noncolinear vectors $\mathbf{u}$ and $\mathbf{v}$ as follows. Define an orthonormal basis \(\{\mathbf{e}_x, \mathbf{e}_y\}\) in the plane orthogonal to \(\mathbf{u}\), adapted to the direction of \(\mathbf{v}\), as
\begin{equation}
\mathbf{e}_x = \frac{(\mathbf{u} \times  \mathbf{v}) \times     \mathbf{u}}{\sqrt{1 - (\mathbf{u} \cdot \mathbf{v})^2}}, \quad
\mathbf{e}_y = \frac{\mathbf{u} \times \mathbf{v}}{\sqrt{1 - (\mathbf{u} \cdot  \mathbf{v})^2}}\,.
\label{eq:basis_1}
\end{equation}
Similarly, we can define, in the plane orthogonal to $\mathbf{v}$, the vectors
\begin{equation}
\mathbf{e}_\theta = -\frac{(\mathbf{v} \times \mathbf{u}) \times \mathbf{v}}{\sqrt{1 - (\mathbf{u} \cdot \mathbf{v})^2}}, \quad
\mathbf{e}_\phi = \frac{\mathbf{u} \times \mathbf{v}}{\sqrt{1 - (\mathbf{u} \cdot \mathbf{v})^2}}\,.
\label{eq:basis_2}
\end{equation}
Using these, the matrix \(U^{ij}(\mathbf{u},   \mathbf{v})\) is given by
\begin{equation}
U^{ij}(\mathbf{u},  \mathbf{v}) = (\mathbf{e}_x)^i (\mathbf{e}_\theta)^j + (\mathbf{e}_y)^i (\mathbf{e}_\phi)^j\,.
\label{eq:gamma_matrix}
\end{equation}
The function $\mathcal{T}(\Theta)$ is shown alongside the Helling-Downs (its PTA counterpart) in Fig.~\ref{fig:hd_Mihaylov}.

%%%%%%%%%%%%%%%%%%%%%%%%%%%%%%%%%%%%%%%%%
%%%%%%%%%%%%%%%%%%%%%%%%%%%%%%%%%%%%%%%%%%

\subsection{Relative astrometry ORF}
 In relative astrometry, we observe the change in the relative angle $\psi^k_{12}=\arccos{(\mathbf{n}_{k1} \cdot \mathbf{n}_{k2})}$ between pairs of stars, labelled by $k=a,b..$, where $(\mathbf{n}_{k1}, \mathbf{n}_{k2})$ are the directions to the stars in the $k$-th pair. From these quantities which we can construct correlators like $\langle \delta\!\cos{\psi^a_{12}}\, \delta\!\cos{\psi^b_{12}} \rangle$. As in the case of absolute astrometry, we can write the correlator as
\begin{equation}
\textstyle
\langle \delta\!\cos{\psi^a_{12}}\, \delta \!\cos{\psi^b_{12}} \rangle \!\propto \!\mathcal{H}_{\scriptstyle \!D\!A}\hspace{-0.07em}(\mathbf{n}_{a_1},\mathbf{n}_{a_2},\mathbf{n}_{b_1},\mathbf{n}_{b_2})\!\times\!S_{h}(f)\,,
\label{eq:cos_corr}
\end{equation}
i.e. in a similar way to Eq.~\eqref{eq:Mihaylov}, where $\mathcal{H}_{DA}$ is the ORF that is now a scalar quantity. The factor $\mathcal{H}_{DA}$ depends on the configuration of the four stars (two pairs) and, in the more general case, is specified by five angles. 

To evaluate $\mathcal{H}_{DA}$ a key observation is that the correlator $\bigl\langle \delta \!\cos\psi^a_{12}\,\delta\!\cos\psi^b_{12} \bigr\rangle$ can be expressed in terms of absolute astrometry (i.e. single star) correlators using Eq.~\eqref{eq:delta_cos}
\begin{align}
\bigl\langle \delta \!&\cos\psi^a_{12}\,\delta\!\cos\psi^b_{12} \bigr\rangle\propto\nonumber\\[1.2ex]
&\bigl\langle \bigl( \mathbf{n}_{a_1} \cdot \delta \mathbf{n}_{a_2} + \delta \mathbf{n}_{a_1} \cdot \mathbf{n}_{a_2} \bigr) 
\bigl( \mathbf{n}_{b_1} \cdot \delta \mathbf{n}_{b_2} + \delta \mathbf{n}_{b_1} \cdot \mathbf{n}_{b_2} \bigr) \bigr\rangle\,,
\nonumber\\[1.2ex]
&=\mathbf{n}_{a_1} \cdot \bigl\langle \delta \mathbf{n}_{a_2}\, \delta \mathbf{n}_{b_2} \bigr\rangle \cdot \mathbf{n}_{b_1} 
\!+ \mathbf{n}_{a_2} \cdot \bigl\langle \delta \mathbf{n}_{a_1}\, \delta \mathbf{n}_{b_2} \bigr\rangle \cdot  \mathbf{n}_{b_1} 
\! \nonumber \\[1.2ex] &+ (b_1\!\leftrightarrow\! b_2)\,.
\label{eq:rel_from_abs_1}
\end{align}
This means by using Eq.~\eqref{eq:Mihaylov} and Eq.~\eqref{eq:Mihaylov_bis} we can write
\begin{align}
\mathcal{H}_{\scriptstyle \!D\!A}\hspace{-0.07em}(\mathbf{n}_{a_1}&,\mathbf{n}_{a_2},\mathbf{n}_{b_1},\mathbf{n}_{b_2})\propto
 n_{a_1}^i \mathcal{T}(\psi_{a_2b_2}) U_{a_2b_2}^{ij} n_{b_1}^j 
\! \nonumber \\
&+n_{a_2}^i \mathcal{T}(\psi_{a_1b_2})U_{a_1b_2}^{ij} n_{b_1}^j 
\!+ \! (b_1\!\leftrightarrow\! b_2)\,,
\label{eq:rel_from_abs_2}
\end{align}

%%%%%%%%%%%%%%%%%%%%%%%%%%%%%%%%%%%%%%%%%
\begin{figure}[thp]
\centering
    \includegraphics[width=0.49\textwidth]{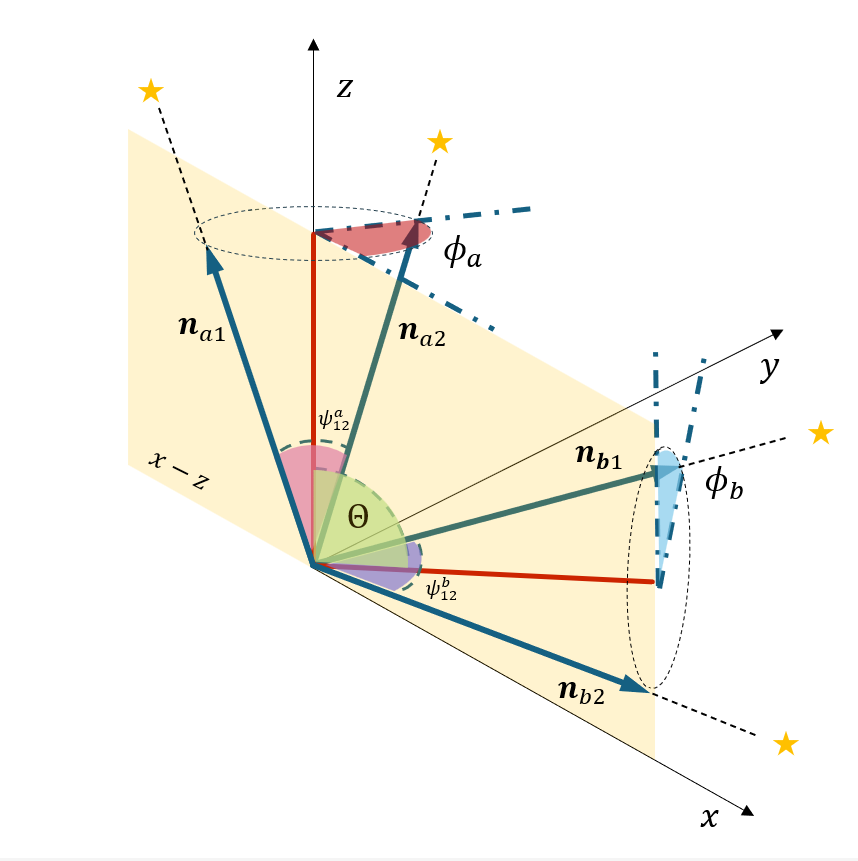}
    \caption{Geometrical configuration of the direction vectors to two pairs of stars, $(\mathbf{n}_{a1},\mathbf{n}_{a2})$ and $(\mathbf{n}_{b1},\mathbf{n}_{b2})$, in the sky parametrized by five angles. We used the freedom in performing a rigid rotation to align the bisector of the directions $(\mathbf{n}_{a_1},\mathbf{n}_{a_2})$ to the stars of the pair $a$, along the $z$ axis, while the bisector of the directions $(\mathbf{n}_{b_1},    \mathbf{n}_{b_2})$ to the stars of the pair $b$, lies in the $x-z$ plane.}
    \label{fig:conf_2}
\end{figure}
%%%%%%%%%%%%%%%%%%%%%%%%%%%%%%%%%%%%%%%%%%
 where $\psi_{{a_i}{b_j}}=\arccos{(\mathbf{n}_{a_i} \cdot \mathbf{n}_{bj})}$ are the angles betwen the $i$-th star in the $a$ pair and the $j$-th star in the $b$ pair.  
 
 A convenient choice for the four vectors is to take the bisector of the first pair aligned with the $z$ axis and the bisector of the second pair to lie in the $x-z$ plane separated by an angle $\Theta$, as displayed in Fig.~\ref{fig:conf_2}. With this choice, the configurations of the stars is completely specified by two angles $\psi^a_{12},\psi^b_{12}$, between the stars in each pair, and three angles $(\Theta, \phi_a, \phi_b)$. The angle $\Theta$ is the angle between the two bisectors of each star pair. The angle $\phi_a$ is measured between the direction connecting the two stars in the $a$-pair, represented by the vectors $\mathbf{n}_{a1}$ and $\mathbf{n}_{a2}$, and the $x$-$z$ plane, which contains the two bisectors. Similarly, $\phi_b$ is the angle between the direction connecting $\mathbf{n}_{b1}$ and $\mathbf{n}_{b2}$ and the $x$-$z$ plane. We can express the directions $(\mathbf{b}_{a_1},\mathbf{b}_{a_2},\mathbf{b}_{b_1},\mathbf{b}_{b_2})$ in terms of $(\psi^a_{12},\psi^b_{12},\Theta, \phi_a, \phi_b)$ as
\begin{align*}
\mathbf{n}_{a_1} &= 
\begin{bmatrix}
\sin\left(\frac{\psi^a_{12}}{2}\right) \cos\phi_a \\
\sin\left(\frac{\psi^a_{12}}{2}\right) \sin\phi_a \\
\cos\left(\frac{\psi^a_{12}}{2}\right)
\end{bmatrix}\,, \,
\mathbf{n}_{a_2} = 
\begin{bmatrix}
\sin\left(\frac{\psi^a_{12}}{2}\right) \cos(\phi_a + \pi) \\
\sin\left(\frac{\psi^a_{12}}{2}\right) \sin(\phi_a + \pi) \\
\cos\left(\frac{\psi^a_{12}}{2}\right)
\end{bmatrix}\,, \\[1ex]
\mathbf{n}_{b_1} &= 
\begin{bmatrix}
\cos\left(\frac{\psi^b_{12}}{2}\right) \sin\Theta - \cos(\phi_b + \pi) \sin\left(\frac{\psi^b_{12}}{2}\right) \cos\Theta \\
\sin\left(\frac{\psi^b_{12}}{2}\right) \sin(\phi_b + \pi) \\
\cos\left(\frac{\psi^b_{12}}{2}\right) \cos\Theta + \cos(\phi_b + \pi) \sin\left(\frac{\psi^b_{12}}{2}\right) \sin\Theta
\end{bmatrix}\,, \\[1ex]
\mathbf{n}_{b_2} &= 
\begin{bmatrix}
\cos\left(\frac{\psi^b_{12}}{2}\right) \sin\Theta - \cos\phi_b\, \sin\left(\frac{\psi^b_{12}}{2}\right) \cos\Theta \\
\sin\left(\frac{\psi^b_{12}}{2}\right) \sin\phi_b \\
\cos\left(\frac{\psi^b_{12}}{2}\right) \cos\Theta + \cos\phi_b\, \sin\left(\frac{\psi^b_{12}}{2}\right) \sin\Theta
\end{bmatrix}\,.
\end{align*}
Using these expressions one can construct the $U_{ij}$ matrices as in Eq.~\eqref{eq:gamma_matrix} and evaluate explicitly $\mathcal{H}_{DA}$ from Eq.~\eqref{eq:rel_from_abs_2}. 

\begin{figure}[thp]
\centering
\includegraphics[width=0.5\textwidth]{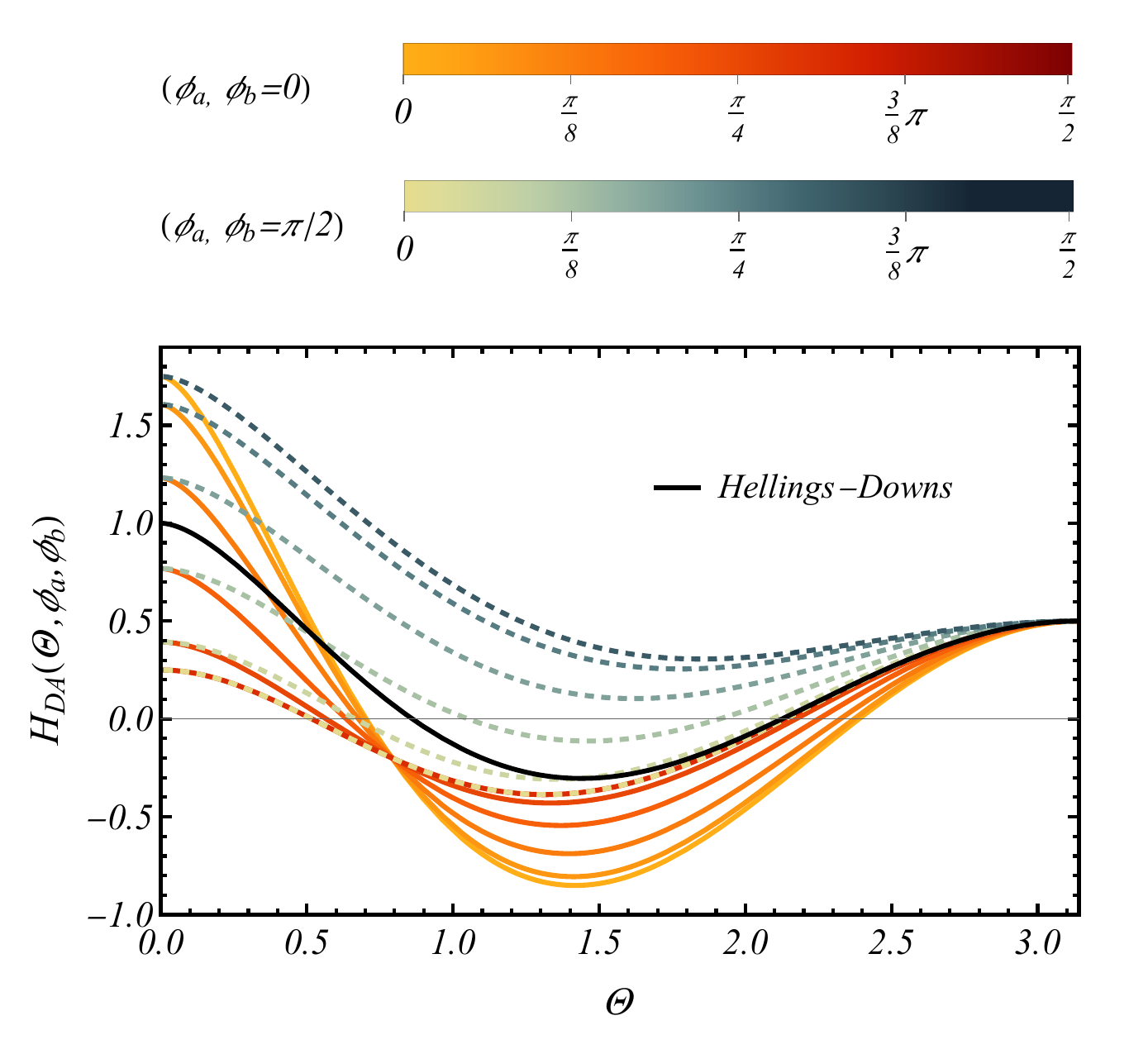}
    \caption{The plot shows the ORF $\mathcal{H}_{DA}(\Theta, \phi_a, \phi_b)$ as a function of $\Theta$ for $\phi_b = 0$ (solid lines) and $\phi_b = \pi/2$ (dashed lines), and for five values of $\phi_a$ in $(0, \pi/2)$, indicated by different colors as specified by the colorbars above. The black solid curve represents the Hellings–Downs correlation, which is also recovered by averaging $\mathcal{H}_{DA}(\Theta, \phi_a, \phi_b)$ over $\phi_a$ and $\phi_b$. 
    }
    \label{fig:hda}
\end{figure}

The picture described above simplifies when the angles between the stars in each pair are small, i.e. $\psi^a_{12}=\psi^b_{12}= \psi\sim 0$, in which case these angles factor out, as $\bigl\langle \delta \!\cos\psi^a_{12}\,\delta\!\cos\psi^b_{12} \bigr\rangle \propto \psi^2 \bigl\langle \delta \!\psi^a_{12}\,\delta\!\psi^b_{12} \bigr\rangle \propto \psi^4 \mathcal{H}_{DA}(\Theta, \phi_a, \phi_b)$. The final expression for $\mathcal{H}_{DA}(\Theta, \phi_a, \phi_b)$ is
\begin{widetext}
\begin{align}
&\mathcal{H}_{\mathcal{DA}}(\Theta, \phi_a, \phi_b) = \frac{\pi}{12 (1 + \cos{\Theta})^2} \Bigg[ 22 + 31 \cos{\Theta} + 10 \cos{(2 \Theta)} + \cos{(3 \Theta)} - 15 \cos{\left( 2 (\phi_a + \phi_b - \Theta) \right)} \nonumber\\ 
&- 15 \cos{\left( 2 (\phi_a + \phi_b + \Theta) \right)}+ 12 \cos{(\Theta - 2 (\phi_a + \phi_b))} + 12 \cos{(\Theta + 2 (\phi_a + \phi_b))} 
+ 24 \log{\sin{\left( \frac{\Theta}{2} \right)}} \nonumber\\
&+ 24 \cos{(\phi_b - \phi_a)} \cos{(\phi_a + \phi_b)} \sin^2{\Theta}+ 6 \Bigg\{ \cos{(2 (\phi_a + \phi_b))} \left( 9 + 4 \left( 11 - 12 \cos{\Theta} \right) \log{\sin{\left( \frac{\Theta}{2} \right)}} \right) \nonumber\\
&+ 4 \cos{(\phi_b - \phi_a)} \cos{(\phi_a + \phi_b)} \left( \cos{\Theta} + 4 \log{\sin{\left( \frac{\Theta}{2} \right)}} \right) \sin^2{\Theta}
+ 2 \log{\sin{\left( \frac{\Theta}{2} \right)}} \left( \cos{\Theta} - \cos{(3 \Theta)} \right. \nonumber \\
&\left.  - 4 \cos{(2 \Theta)} \sin^2{(\phi_a + \phi_b)} \right) \Bigg\} \Bigg]\,,
\label{eq:hda}
\end{align}
\end{widetext}
and is shown in Fig.~\ref{fig:hda} as a function of $\Theta$ and for different values of $\phi_a$ and $\phi_b$. Expression ~\eqref{eq:hda} exibiths symmetry under the exchange $(\phi_a \leftrightarrow \phi_b)$ and for simultaneous change $(\phi_a \to -\phi_a, \phi_b \to -\phi_b)$. Furthermore, eemarkably, the average of Eq.~\eqref{eq:hda} over the angles $\phi_a$ and $\phi_b$ yields the Hellings--Downs curve
\[
\frac{1}{(2\pi)^2} 
\int_0^{2\pi} \int_0^{2\pi} \mathcal{H}_{DA}(\Theta,\phi_a,\phi_b)\,d\phi_a\,d\phi_b = \mathcal{H}(\Theta)\,,
\]
where $\mathcal{H}(\Theta)\propto\int d\Omega_p\, \langle \delta \nu(\mathbf{n},\mathbf{p}) \delta \nu(\mathbf{m},\mathbf{p})\rangle$, $\delta \nu$ being the GW-induced redshift in the time pulses, $\mathbf{n}$ and $\mathbf{m}$ the directions to the pulsars and $\Theta=\arccos{\mathbf{n}\cdot \mathbf{m}}$.  

%%%%%%%%%%%%%%%%%%%%%%%%%%%%%%%%%%%%%%%%%
%%%%%%%%%%%%%%%%%%%%%%%%%%%%%%%%%%%%%%%%%%

\subsection{Astrometry–PTA Cross-Correlation ORF}
\begin{figure}[thp]
\centering
\includegraphics[width=0.5\textwidth]{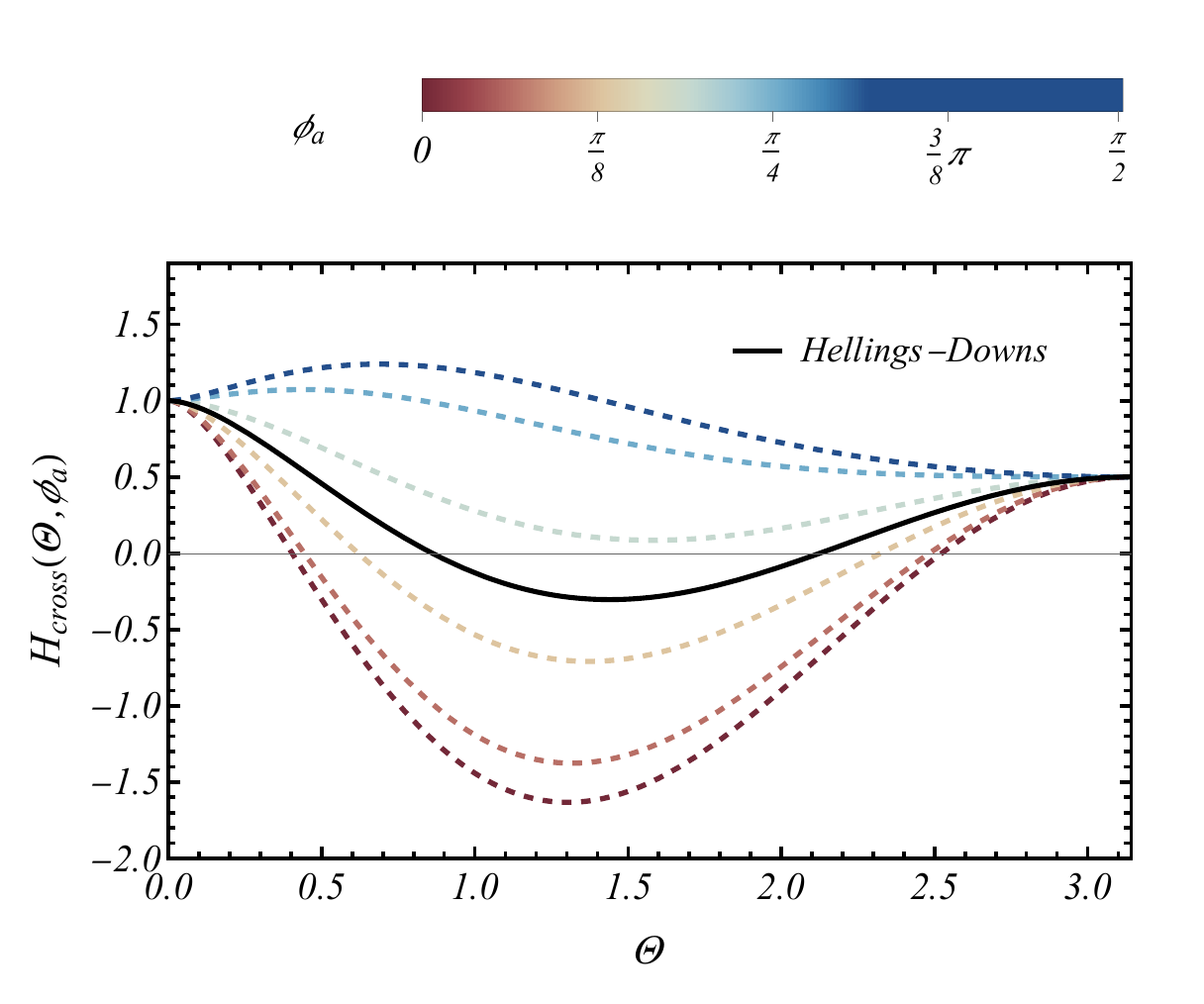}
    \caption{The plot shows the cross-correlation ORF $\mathcal{H}_\text{cross}(\Theta, \phi_a)$ between GW-induced redshift and astrometric deflection as a function of the pulsar-pair angular separation $\Theta$. Results are shown for five different values of $\phi_a$ in the range $(0, \pi/2)$, distinguished by color as indicated by the colorbars above. The black solid curve corresponds to the standard Hellings–Downs correlation, which is recovered by averaging $\mathcal{H}_\text{cross}(\Theta, \phi_a)$ over the angle $\phi_a$.
}
    \label{fig:cross-correlation}
\end{figure}

As we will discuss in Sec.~\ref{sec:Fisher_forecast}, astrometric surveys can be sensitive enough to reveal a SGWB on their own. However, they also offer the possibility to search for the signal in the cross-correlation with PTA observations. Similarly to the redshift and astrometric correlation patterns, there is a characteristic dependence of the correlation between a GW-induced pulsar redshift and astrometric deflection in the light from a star, on the separation angle in the sky. Following again Ref.~\cite{Mihaylov:2018uqm} we can express
\begin{equation}
\langle \delta \nu \, \delta n_a^i\rangle \propto \mathcal{H}_{\text{cross}}^{i}(\mathbf{n}_I,\mathbf{n}_a)\times S_h(f)\,,
\end{equation}
where $\mathbf{n}_I$ is the direction of the $I$-th pulsar and $\mathbf{n}_a$ is the direction of the $a$-th star.  The ORF $\mathcal{H}^{i\,\text{cross}}_A(\mathbf{n}_I,\mathbf{n}_a)$ can be expressed in terms of a scalar function as $\mathcal{H}^{i\,\text{cross}}_{A}\propto \mathcal{P}(\Theta)\mathbf{e}_\theta^i$, where $\Theta=\arccos{(\mathbf{n}_I\cdot \mathbf{n}_a)}$ and $\mathbf{e}_\theta^i$ is defined in Eq.~\eqref{eq:basis_2}. For the cross-correlation in relative astrometry, we have to evaluate the correlator
\[
\begin{aligned}
\langle \delta \nu \, \delta \cos{\psi^a_{12}}\rangle= &\langle \delta \nu \,\, (\mathbf{n}_{a_1}\cdot\delta \mathbf{n}_{a_2} + \delta \mathbf{n}_{a_1}\cdot \mathbf{n}_{a_2} )\rangle \nonumber \\
=&\mathbf{n}_{a_1} \langle z \,\cdot\delta \mathbf{n}_{a_2}\rangle + \langle z \, \delta \mathbf{n}_{a_1}\rangle \mathbf{n}_{a_2}\,.
\end{aligned}
\]
Then, defining $\langle \delta \nu \, \delta \cos{\psi^a_{12}}\rangle \propto \mathcal{H}^\text{cross}_{DA}\times S_h(f)$, one can evaluate $\mathcal{H}^\text{cross}_{DA}$ as
\begin{equation}
\mathcal{H}^\text{cross}_{DA}=n_{a_1}^i\mathcal{P}(\psi_{Ia_2})\mathbf{e}^i_{\theta I{a_2}}+n_{a_2}^i\mathcal{P}(\psi_{Ia_1})\mathbf{e}_{\theta I a_1}^i\,.
\end{equation}
As in the previous section, we introduce, for each stellar pair, the angle $\phi_a$, which quantifies the rotation of the pair about the bisector of their directions and, by convention, $\phi_a = 0$ when the two stars lie in the same plane as the pulsar. In the limit when $\psi^a_{12}$ is small, $\phi_a$ is the only additional angle on which $\mathcal{H}^\text{cross}_{DA}$ depends on besides the pulsar-pair angle $\Theta$. The function $\mathcal{H}^\text{cross}_{DA}(\Theta, \phi_a)$ is plotted in Fig.~\ref{fig:cross-correlation}. Just like $\mathcal{H}_{DA}(\Theta,\phi_a,\phi_b)$, $\mathcal{H}^\text{cross}_{DA}(\Theta,\phi_a)$ gives the Hellings-Downs curve as a function of $\Theta$, after averaging over $\phi_a$
\[
\frac{1}{(2\pi)} 
\int_0^{2\pi} \mathcal{H}^\text{cross}_{DA}(\Theta,\phi_a)\,d\phi_a= \mathcal{H}(\Theta)\,.
\]

%%%%%%%%%%%%%%%%%%%%%%%%%%%%%%%%%%%%%%%%%
%%%%%%%%%%%%%%%%%%%%%%%%%%%%%%%%%%%%%%%%%%

\section{Fisher forecasts}
\label{sec:Fisher_forecast}

%In light of the above, 
In this section, we describe the main ingredients that we used to develop a script to forecast the sensitivity of GW astrometry techniques to SGWB, using the Fisher matrix formalism. The code is based on \texttt{fastpta}\footnote{\url{https://github.com/Mauropieroni/fastPTA}}~\cite{forecasting_pta_gwb} and is extended to include absolute and relative astrometric response as well as cross-correlations between PTA and Gaia. 

\subsection{Noise models and sensitivity}

Estimating the sensitivity to a given signal requires the assumption of a noise model. For each object $I$ considered (PTA or astrometry time residual), we have a set of observations that can be written in the frequency domain as $\tilde d_I (f) =\tilde n_I (f) +\tilde h_I (f)$ 
with Gaussian, stationary noise $\tilde n_I$ and the GW signal $\tilde h_I$. In this work, we consider both pulsars and stars and write their noise PSD in characteristic strain unit $\langle \tilde h^2(f) \rangle = f S_h(f)$. The pulsar noise $\tilde n_a(f)$ statistics for the pulsar $a=1...N_{\rm PTA}$ in the frequency domain follows the statistics
\begin{align}
\langle \tilde n_{a}(f)\tilde n_{b}(f') \rangle &= 12 \pi^2 f^2 \left[ 2\sigma_{P}^2 \Delta t_{P} + \frac{\mathcal{A}^2\left(\frac{f_{yr}}{f} \right)^{\gamma}}{12 \pi^2 f^3_{yr}}  \right]\delta_{ff'}\delta_{ab}\,,\nonumber\\
&\equiv S_{n,P}(f)\delta_{ff'}\delta_{ab},
\label{eq:pulsar_noise}\,,
\end{align}
where, for the sake of simplicity, we assume that all the pulsars have the same noise properties. Here $\Delta t_P$ represents the effective cadence of the PTA survey, $\sigma_P$ the white noise budget error in each of the time residuals, while $\mathcal{A}$, $f_{\rm yr}$ and $\gamma$ parameterize the red noise spectrum, a low frequency noise with a power-law spectrum. In the literature, it is often referred to as spin-noise, a time-correlated noise arising from stochastic angular momentum exchanges between the solid crust and superfluid interior of pulsars~\cite{NANOGrav:2023ctt, EPTA:2023akd}.
In reality, pulsars have very different noise properties \cite{NANOGrav:2023ctt, EPTA:2023akd}. This can produce an uneven response to GWs across the sky because some pulsars contribute significantly more to the total SNR~\cite{quality_quantity}. Here, we have chosen noise parameters such that the SNR of the GWB injected in the PTA is between $3$ and $5$, consistent with the current status of PTA observations as reported in ~\cite{NANOGrav:2023gor, EPTA:2023fyk, Reardon_2023}. The astrometric noise $\tilde n_I(f)$ for the star $I=1...N_{\rm Stars}$ in the time domain follows the statistics
\begin{align}
\langle \tilde n_{m}(f)\tilde n_{n}(f') \rangle &= 2\sigma_{G}^2 \Delta t_{G} \delta_{ff'}\delta_{mn}\,,\nonumber\\
&\equiv S_{n,S}(f)\delta_{ff'}\delta_{mn}\,,
\label{eq:star_noise}
\end{align}
where $\sigma^2_{G}$ is the measurement noise level in the time residual for each star.

Up to now, the Gaia mission has observed more than $10^8$ stars~\cite{vallenari2023gaia}. A challenge of Gaia data analysis for GW detection is to analyse all available data at once, making it computationally expensive due to a very high-rank matrix inversion. This is also true when performing a Fisher forecast.
In~\cite{Moore:2017ity}, a workaround is presented where they pixelize the sky and consider effective astrometric displacements in each pixel, allowing efficient data compression and considerable computational saving, at the cost of a small impact on the overall sensitivity.

The scheme proposes to average the time-dependent astrometric displacements of all the stars present in each of the $N_{\rm pix}$ sky pixels, in such a way that the error on the averaged measurement $\sigma_{\rm pix}$ is mitigated
\begin{align}\label{eq:pixelized_dn}
\delta \hat n_{\rm pix}(t)&=\frac{1}{N_{\rm spp}}\sum_{i=1}^{N_{\rm spp}}\delta \hat n_{i}(t)\,,\nonumber\\
\sigma_{\rm pix}&=\frac{\sigma}{\sqrt{N_{\rm spp}}}\,,
\end{align}
where the number of stars-per-pixel $N_{\rm spp}=N_{\rm Stars}/N_{\rm pix}$ is assumed to be roughly the same for all the pixels considered.

We choose to express all noise components in characteristic strain because different datasets have different units\footnote{The prefactors in powers of $f$ of Eq.~\eqref{eq:pulsar_noise} and Eq.~\eqref{eq:star_noise} serve the purpose of converting the noise PSD in to characteristic strain units.}. PTAs work with timing residuals, the integral over time of the experienced redshift, that have units of time~\cite{NANOGrav:2023ctt, EPTA:2023akd}. The astrometric displacement observed by Gaia is instead directly proportional to the GW strain~\cite{Mihaylov:2018uqm}. Ensuring that the noise PSDs of different detectors are expressed in the same unit is necessary to perform cross-detector correlation calculations.

Additionally, in pulsar timing experiments, a quadratic fit $\sim t^2$ is performed in time domain to account for the deceleration of the pulsar spin rate~\cite{EPTA:2023sfo}. For Gaia data, a similar fit accounts for the proper motion of stars on the celestial sphere~\cite{lindegren2021gaia}. In the frequency domain, it approximately translates into a transmission function $\mathfrak{T}_I(f)$ that filters out lower frequencies (behaving as a high-pass filter $\sim f^6$) as demonstrated in Appendix~\ref{app:transmission_function}. This step is essential to remove any possible degeneracies between very low frequency features and the actual GW signal. For a signal $h_I$ observed in detector $I$, at frequencies close to $1/T_{\rm obs,I}$ the filter behaves as
\begin{equation}
    \mathfrak{T}_I(f) \approx 1 - 1/(f T_{\textrm{obs},I})^6\,.
\end{equation}
The full expression of $\mathfrak{T}_I(f)$ is given in appendix in Eq.~\eqref{eq:transmission_fourier} and Eq.~\eqref{eq:psi_fourier}. Given that a GWB signal is present in the data with characteristic strain $\langle h_{GWB}^2 \rangle = f S_h (f)$, we can write the signal covariance matrix as
\begin{equation}
    C_{h,IJ} = \langle \tilde{h}_I \tilde{h}_J \rangle = f S_h (f) \Gamma_{IJ} R_{IJ}\,,
\end{equation}
where we defined $R_{IJ} =  \left [ \mathfrak{T}_I(f) \mathfrak{T}_J(f) T_{IJ} / T_{\textrm{obs}} \right ]^{1/2}$, the response tensor, $\Gamma_{IJ}$ the ORF introduced in Sec.~\ref{sec:response_to_sgwb} and $T_{IJ}=\textrm{min}[T_{\textrm{obs},I}, T_{\textrm{obs},J}]$ the term accounting for the minimum overlapping time between detectors $I$ and $J$. The $\mathfrak{T}_I(f)$ removes the low frequency content of the signal close to the lowest frequency bin $1/T_{\textrm{obs}}$.

We can now write the covariance matrix
\begin{equation}
    C_{IJ} = \langle \tilde{d}_I \tilde{d}_J \rangle = \delta_{IJ} \langle \tilde{n}^2_I \rangle + C_{h,IJ}\,.
\label{eq:full_covariance}
\end{equation}
This is the central quantity in the forecast since it encapsulates the noise and signal parameters for every combined detector. Finally, we need to define two quantities, the SNR and the Fisher information matrix.\\

\paragraph{Signal to noise ratio and sensitivity.}

Following~\cite{forecasting_pta_gwb}, the SNR is defined as the ratio between the signal covariance and the total covariance, and the sum is performed over all possible combinations of indices.
\begin{equation}\label{eq:SNR_def}
\begin{aligned}
    \textrm{SNR}^2 & = \sum_{f_k} C_{IJ}^{-1} C_{h,JK} C_{KL}^{-1} C_{h,LI}\,,\\
    & = \sum_{f_k} \left[ C_{IJ}^{-1} \Gamma_{JK} R_{JK} C_{KL}^{-1} \Gamma_{LI} R_{LI} \right] f^2 S_h^2\,,\\
    & = \sum_{f_k} \left(\frac{S_h}{S_{\textrm{eff}}}\right)^2\,.
\end{aligned}
\end{equation}
where
\begin{equation}
    f S_{\textrm{eff}} (f) = \left [ C_{IJ}^{-1} C_{KL}^{-1} \Gamma_{JK} R_{JK} \Gamma_{LI} R_{LI} \right]^{-1/2},
\end{equation}
giving the sensitivity curve associated with the fully combined detector and its noise properties\footnote{In the weak signal limit where $\langle h^2 \rangle \ll \langle n^2_I \rangle$, one can show that the effective noise $S_{\textrm{eff}} (f)$ takes the form presented in~\cite{Hazboun_2019} $S_{\textrm{eff}} (f) \approx \left [ \sum_{I \neq J} \frac{T_{IJ}}{T_{\textrm{obs}}} \frac{\mathfrak{T}_I(f) \mathfrak{T}_J(f) \Gamma_{IJ}^2}{\langle \tilde{n}_I^2 \rangle \langle \tilde{n}_J^2 \rangle}\right]^{-1/2}$}. \\

\paragraph{Fisher information matrix.}

The Fisher information matrix for Gaussian and stationary noise is given by~\cite{forecasting_pta_gwb}

\begin{equation}
\begin{aligned}
    F_{\alpha \beta} & = \sum_{f_k} \textrm{tr} \left \{ C \frac{\partial C}{\partial \theta_\alpha} C \frac{\partial C}{\partial \theta_\beta} \right \},\\
    & = \sum_{f_k} C_{IJ} \frac{\partial R_{JK} S_h}{\partial \theta_\alpha} C_{KL} \frac{\partial R_{LI} S_h}{\partial \theta_\beta},
\end{aligned}
\end{equation}
where $\partial / \partial \theta_\alpha$ the partial derivatives with respect to the signal parameters $\theta_\alpha$. In this work, the partial derivatives are performed with respect to the log amplitude $\log_{10}A_{\rm GW}$ and spectral index $\alpha$ for the signal PSD that follows a power-law spectrum.

\begin{table}
    \centering
    \begin{tabular}{c|c}
    $\sigma_{P}$ & $1$ [$\mu s$] \\
    $\Delta t_{P}$ & $3$ [days]\\
    $\log_{10}\mathcal{A}$ & $-13.7$ \\ 
    $\gamma$ & $3$ \\
    $\sigma_{G}$ & $0.2$ [mas]\\
    $\Delta t_{G}$ & $25$ [days]\\
    $T_{\rm obs}$ & $10$ [years]\\
    $N_{\rm PTA}$ & $25$ \\
    $N_{\rm Stars}$ & $10^8$ \\
    \end{tabular}
    \caption{Noise parameter values used for the forecast.}
    \label{tab:noise_values}
\end{table}

\subsection{Combining datasets}

Forecasting the performance of a combined PTA+Gaia dataset requires an expression of the full covariance matrix in Eq.~\eqref{eq:full_covariance}. Using the noise models of Eq.~\eqref{eq:pulsar_noise} and Eq.~\eqref{eq:star_noise}, we want to assess the ability of a PTA+Gaia dataset to detect a SGWB with a power-law spectrum

\begin{align}\label{eq:signal_PSD}
S_h(f)&=f^{-1}h_{c}^2(f)\,,\nonumber\\
h_{c}(f)&=A_{\rm GW}\left(\frac{f}{f_{\rm ref}}\right)^{\alpha}\,,
\end{align}
where $A_{\rm GW}$ is the signal amplitude and $\alpha$ the spectral index. For the forecast, we compare two sets of values (i) $A_{\rm GW}=10^{-14}$ at $f_{\rm ref}=30$ nHz and $\alpha=0$, corresponding to the values that were recently inferred by the PTA collaborations from their data \cite{EPTA:2023fyk, NANOGrav:2023gor}, (ii) $A_{\rm GW}=3\times 10^{-15}$ at $f_{\rm ref}=30$ nHz and $\alpha=-2/3$ that are the expected parameters for a population of SMBHBs in circular orbit in the Universe \cite{Sesana:2008mz, Phinney:2001di}.

In Sec.~\ref{sec:response_to_sgwb}, the ORFs corresponding to the single direction astrometry in the presence of an isotropic SGWB are calculated. The displacement vector $\delta n$ induced by the GW signal is decomposed on the 2d-plane ($e_\theta, e_\phi$) perpendicular to the unperturbed star position, providing a 2d vector ($\delta \theta,\delta \phi$) that can be directly related to the observed star position. Therefore, the single direction astrometry dataset is two-dimensional and the ORFs are 2$\times$2 matrices. The relative astrometry tracks the evolution of the relative angle between two stars. Similarly to PTA, it is represented by a one-dimensional time series.

The covariance matrix can be written in a block structure where the diagonal contains the information about single detectors (PTA or Gaia) and the off-diagonal contains the cross-detector correlations as
\begin{equation}
C^{P+G} _{IJ}= \begin{bmatrix}
C^{PP} & C^{PG} \\
C^{GP} & C^{GG}
\end{bmatrix}\,,
\end{equation}
with the Pulsar-Pulsar covariance
\begin{equation}
    C^{PP}_{ab} = \delta_{ab} \langle \tilde{n}^2_{PSR,a} \rangle +  f S_h (f) \Gamma_{ab,PP} R_{ab}\,,
\end{equation}
the Gaia-Gaia covariance
\begin{equation}
    C^{GG}_{mn} = \delta_{mn} \langle \tilde{n}^2_{star,n} \rangle +  f S_h (f) \Gamma_{mn,GG} R_{mn}\,,
\end{equation}
and the Pulsar-Gaia covariance
\begin{equation}
    C^{PG}_{am} = C^{GP}_{ma} = f S_h (f) \Gamma_{am,PG} R_{am}\,.
\end{equation}

\subsection{Fast analytical SNR estimates}

\begin{figure}[thp]
\centering
\includegraphics[width=0.5\textwidth]{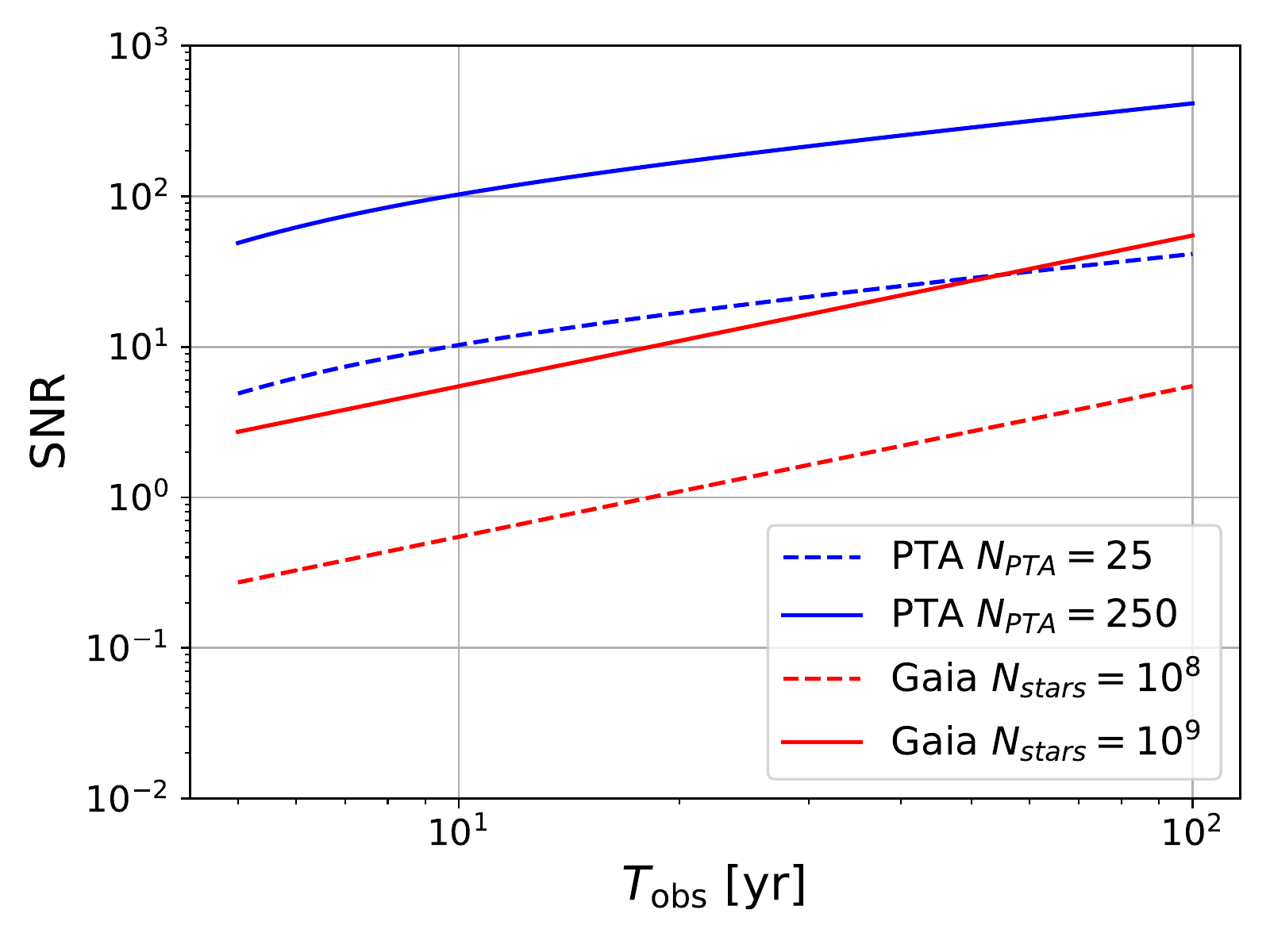}
    \caption{Signal to Noise Ratio as a function of the total time of observation for a PTA and an astrometry (Gaia-like) experiment. The curves are drawn following Eq.~\eqref{eq:SNRs_quick_def} where the noise values for PTA and the astrometric surveys are specified in Tab.~\ref{tab:noise_values}, while the signal power spectrum is the one in of Eq.~\eqref{eq:signal_PSD}, with $A_{\rm GW}=10^{-14}$ and $\alpha=0$.}
    \label{fig:SNRs_quick}
\end{figure}

A quick way to forecast the observability of an SGWB is to analytically approximate Eq.~\eqref{eq:SNR_def} up to its order of magnitude. As explained before, we compress the data of all the astrometric deflections into a set of sky pixels $N_{\rm pix}$, as in Eq.~\eqref{eq:pixelized_dn}.
With this scheme, we can build $N_{\rm pix}(N_{\rm pix}+1)/2\simeq N_{\rm pix}^2/2$ correlators between all the sky pixels and under the assumption that the pixels and the frequencies have uncorrelated noise, we can approximate the SNR of our PTA, astrometry and combined dataset. Furthermore, we approximate the value of the overlap functions $\Gamma_{IJ, PP}$, $\Gamma_{IJ, GG}$ and $\Gamma_{IJ, PG}$ to unity in the SNR formula because the sum of the squared terms roughly equates to their average over all the possible separation angles
\begin{align}\label{eq:SNRs_quick_def}
&\text{SNR}^2_{\rm PTA}\simeq\frac{N^2_{\rm PTA}}{2} T_{\rm obs}\times 2\int_{f_{\rm min}}^{f_{\rm max}} df \frac{S_h^2(f)}{(S_{n,P}(f)+S_h(f))^2}\,,\nonumber\\
&\text{SNR}^2_{\rm Stars}\simeq\frac{N^2_{\rm pix}}{2} T_{\rm obs}\times 2\int_{f_{\rm min}}^{f_{\rm max}} df \frac{S_h^2(f)}{(S_{n,pix}(f)+S_h(f))^2}\,,\nonumber\\
&\text{SNR}^2_{\rm cross}\simeq 2 N_{\rm PTA} N_{\rm pix} T_{\rm obs}\times\int_{f_{\rm min}}^{f_{\rm max}} df S_h^2(f)\times\nonumber\\
&\left[S_{n,P}(f)S_{n,pix}(f)+(S_{n,P}(f)+S_{n,pix}(f))S_h(f)+S_h^2(f)\right]^{-1}\,,\nonumber\\
&\text{SNR}^2_{\rm tot}=\text{SNR}^2_{\rm PTA}+\text{SNR}^2_{\rm Stars}+\text{SNR}^2_{\rm cross}\,,
\end{align}
where $S_{n,pix}(f)=S_{n,S}(f)/N_{\rm spp}$ is the pixel noise PSD, defined via Eq.~\eqref{eq:star_noise} and Eq.~\eqref{eq:pixelized_dn}.
Concerning the minimum frequency that we can exploit, in principle we can set $f_{\rm min}=1/T_{\rm obs}$. However, we expect that similarly to what happens in the PTA case, the subtraction procedure of the fitting methods in the stars' proper motion datastreams will introduce a strong noise at low frequency, effectively cutting down the signal at frequencies below $\sim 2/T_{\rm obs}$~\cite{forecasting_pta_gwb}. The maximum frequency is the Nyquist frequency set by our sampling rate $\sim 2/\Delta T_{P/G}$, and
where we assumed $\gamma=3$ as stated in Tab.~\ref{tab:noise_values}. In Fig. \ref{fig:SNRs_quick}, we show how the SNR varies for different configurations of stars and pulsars as a function of the total time of observation.
Given the best fit value of the measurement by PTA collaborations~\cite{NANOGrav:2023hfp, EPTA:2023fyk, mpta_gwb, Reardon_2023}, we find that $S_{n,pix}(f)\gg S_{h}(f)$ for $f\gtrsim1$nHz, therefore leaving the denominators of $\text{SNR}_{\rm Stars}$ and $\text{SNR}_{\rm cross}$ in Eq.~\eqref{eq:SNRs_quick_def} dominated by the noise term even for larger times of observation $T_{\rm obs}$\footnote{Which means that the Gaia datastreams for stars and pixels are individually noise dominated}.

It is important to note that the astrometric SNR will be the same as that obtained without compressing the dataset into pixels. In fact, the SNR is the same as that obtained when considering all the $N_{\rm tot}$ stars with individual astrometric error $\sigma$, without employing the pixelization method, which immediately follows from Eq.~\eqref{eq:pixelized_dn}.

\section{Results}
\begin{figure}
    \centering
    \includegraphics[width=1.0\linewidth]{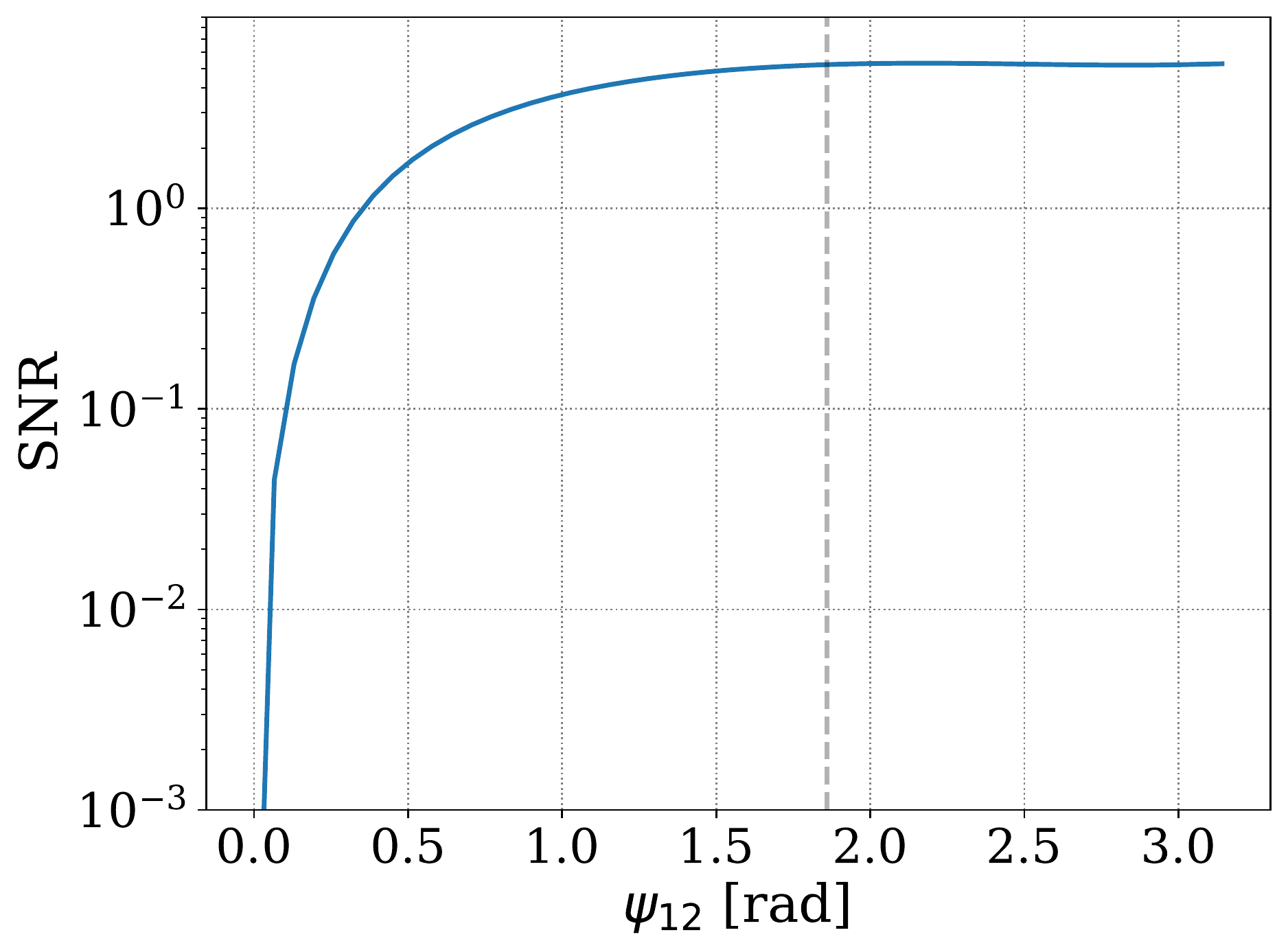}
    \caption{SGWB SNR for relative astrometry as a function of the unperturbed relative angle $\psi_{12}$ for a dataset made of $10^9$ star pairs. The dashed line shows where $\psi_{12}=106.5^\circ$. For small angles, the SNR goes to zero. The SNR is calculated using Eq.~\eqref{eq:SNR_def}, considering a SGWB with $A_{\rm GW}=10^{-14}$ and $\alpha=0$.}
    \label{fig:SNR_psi}
\end{figure}

\begin{table*}[]
    \centering
    \begin{tabular}{c|cc|cc}
        & $A_{\rm GW}=3\times 10^{-15}$ & ; $\alpha=-2/3$ & $A_{\rm GW}=10^{-14}$ & ; $\alpha=0$ \\ \hline
        $N_{\rm stars}$ & $10^8$ & $10^9$ & $10^8$ & $10^9$ \\
        \hline
        PTA & 3.2 & 3.2 & 5.2 & 5.2 \\
        PTA+Gaia & 3.3 & 4.1 & 5.3 & 6.7 \\
        PTA + R-Gaia & 3.3 & 4.4 & 5.3 & 7.3 
    \end{tabular}
    \caption{SGWB SNR calculated for different combined datasets and properties of the SGWB using Eq.~\eqref{eq:SNR_def}.}
    \label{tab:SNR_table}
\end{table*}

%%%%%%%%%%%%%%%%%%%%%%%%%%%%%%%%%%%%%%%%%
\begin{figure*}
\centering
    \includegraphics[width=\linewidth]{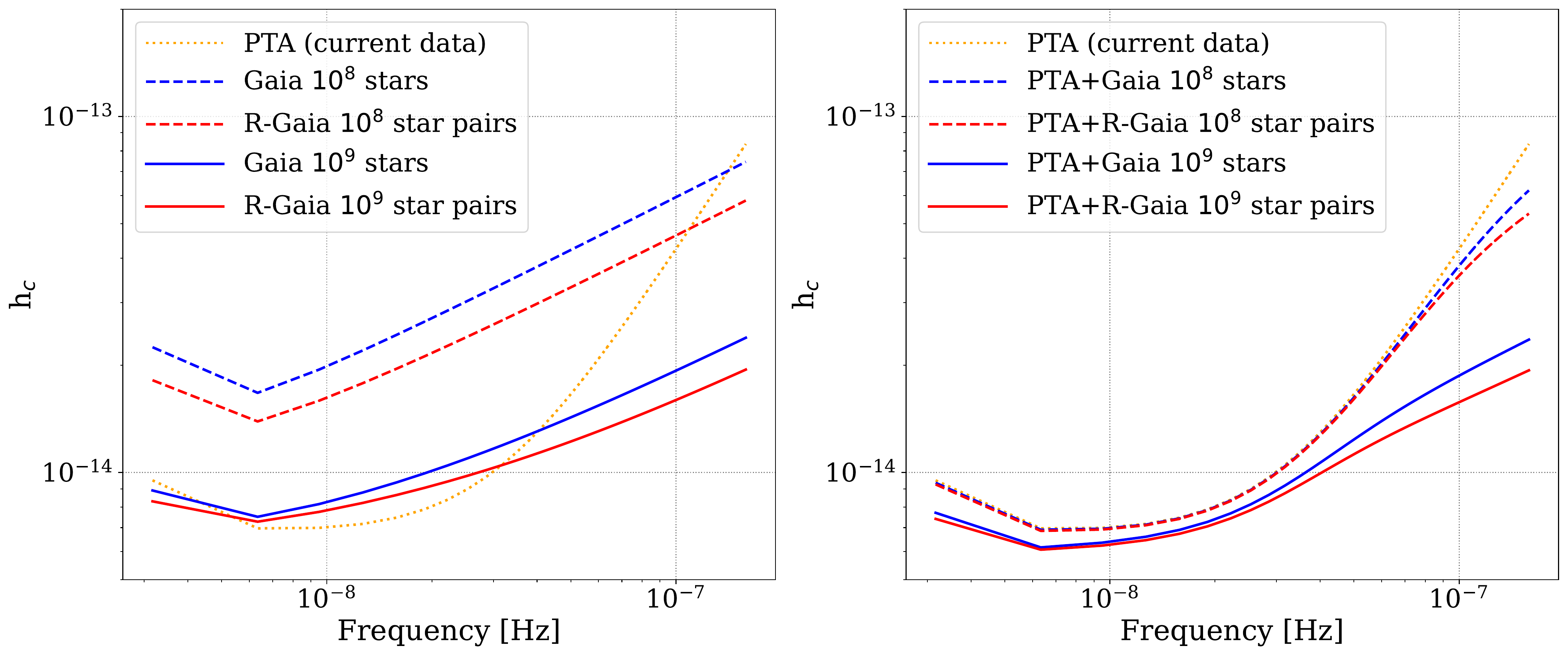}
    \caption{(left) The sensitivity curves evaluated at the harmonics of $1/T_{\rm obs}$  for single-star astrometry (Gaia) and relative star astrometry (R-Gaia) with unperturbed relative angle of $106.5^\circ$ for $N_{\rm stars}=10^8$ and $N_{\rm stars}=10^9$ using the noise parameter values of Tab.~\ref{tab:noise_values} (right) The sensitivity curves evaluated at the harmonics of $1/T_{\rm obs}$ for the combined datasets PTA+single-star astrometry (PTA+Gaia) and PTA+relative star astrometry (PTA+R-Gaia) for $N_{\rm stars}=10^8$ and $N_{\rm stars}=10^9$ with unperturbed relative angle of $106.5^\circ$ using the noise parameter values of Tab.~\ref{tab:noise_values}. The considered SGWB has a power-law spectrum with $A_{\rm GW}=10^{-14}$ and $\alpha=0$.
    }
    \label{fig:sensitivity curves}
\end{figure*}
%%%%%%%%%%%%%%%%%%%%%%%%%%%%%%%%%%%%%%%%%%

\begin{figure*}[t]
  \centering
  \begin{minipage}[t]{0.49\textwidth}
    \centering
    \includegraphics[width=\linewidth]{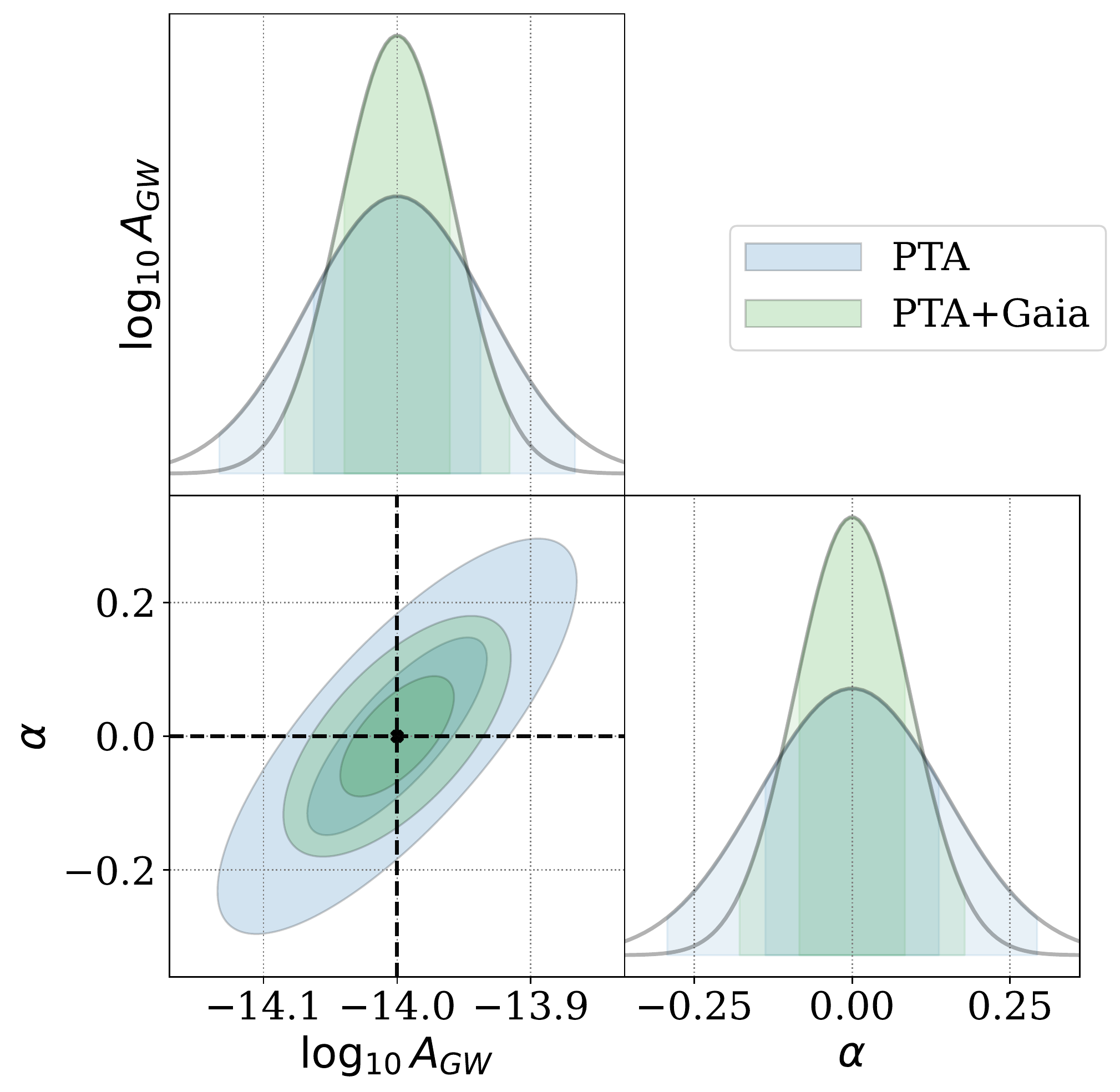}
  \end{minipage}
  \hfill
  \begin{minipage}[t]{0.49\textwidth}
    \centering
    \includegraphics[width=\linewidth]{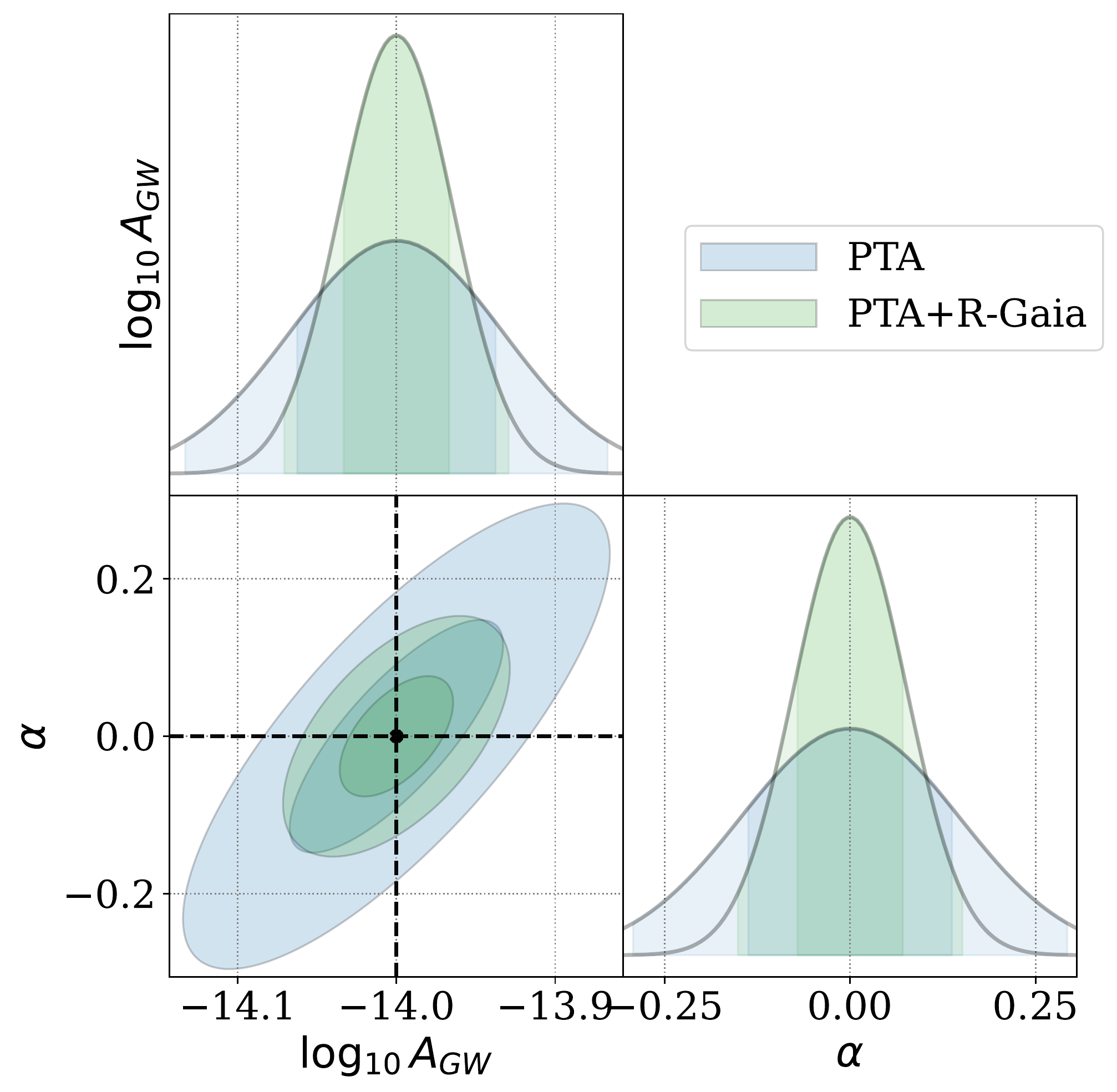}
  \end{minipage}
  \caption{Uncertainties on the amplitude $A_{\rm GW}$ and spectral index $\alpha$ of a SGWB with a power-law spectrum with $A_{\rm GW}=10^{-14}$ and $\alpha=0$, detected using the combined PTA+Gaia dataset (left) and the PTA+R-Gaia dataset (right). The shaded areas represent the $1$ and $2\sigma$ credible regions.}
  \label{fig:fisher_ptagaia}
\end{figure*}

\begin{figure*}[t]
  \centering
  \begin{minipage}[t]{0.49\textwidth}
    \centering
    \includegraphics[width=\linewidth]{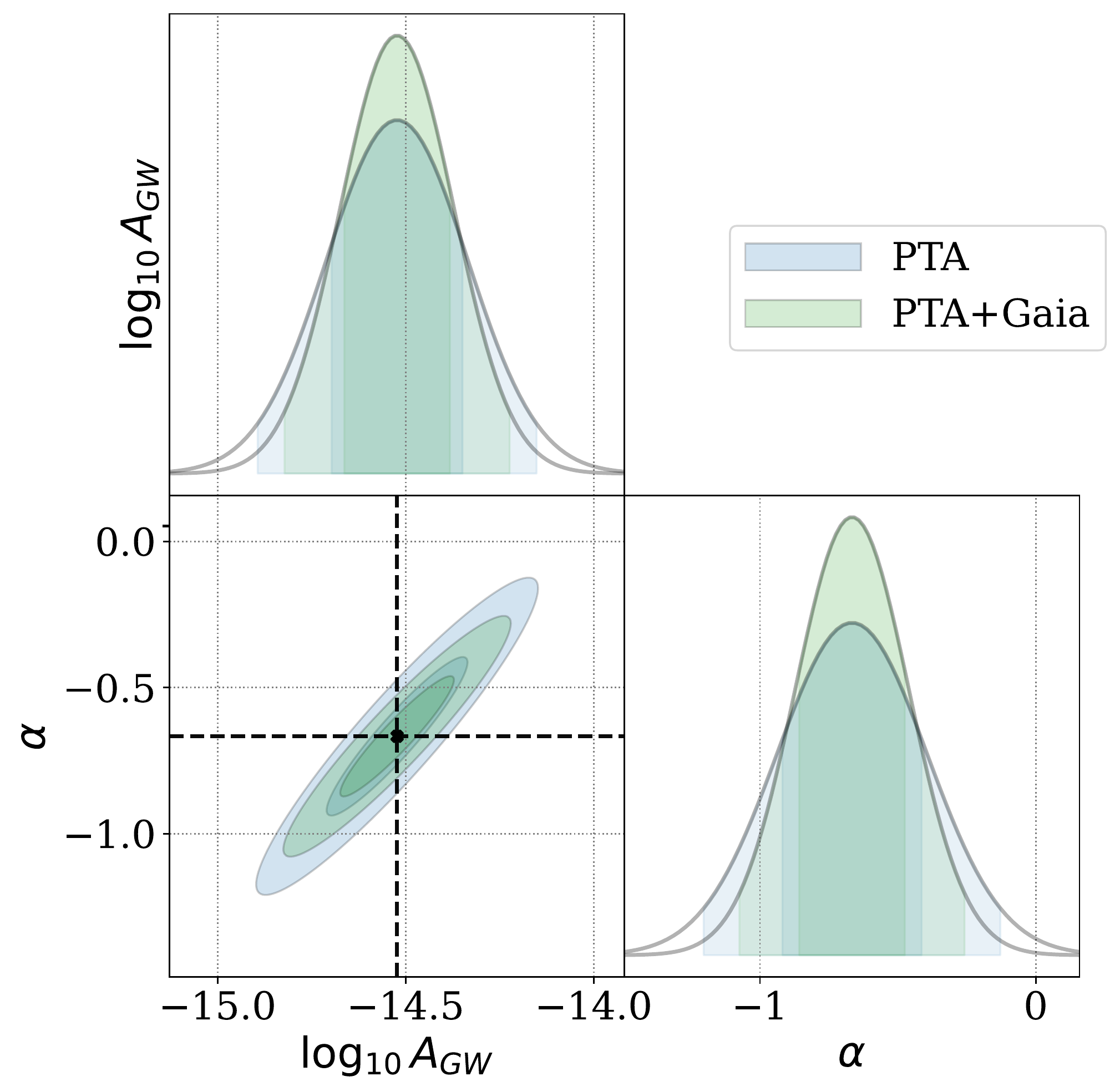}
  \end{minipage}
  \hfill
  \begin{minipage}[t]{0.49\textwidth}
    \centering
    \includegraphics[width=\linewidth]{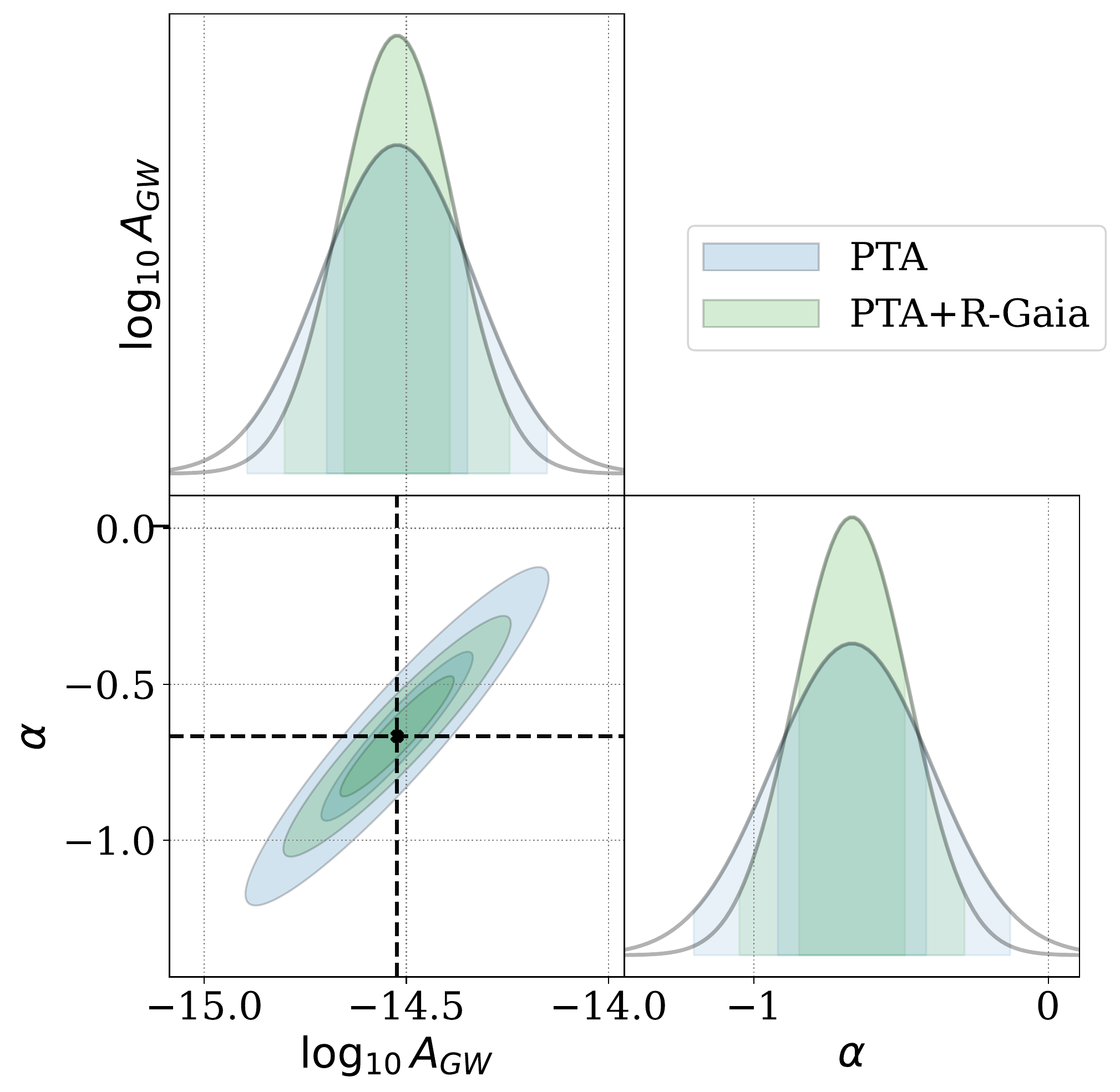}
  \end{minipage}
  \caption{Uncertainties on the amplitude $A_{\rm GW}$ and spectral index $\alpha$ of a SGWB with a power-law spectrum with $A_{\rm GW}=3 \times 10^{-15}$ and $\alpha=-2/3$, detected using the combined PTA+Gaia dataset (left) and the PTA+R-Gaia dataset (right). The shaded areas represent the $1$ and $2\sigma$ credible regions.}
  \label{fig:fisher_ptagaia_astro}
\end{figure*}

\label{sec:results}
Using the noise parameter values presented in Tab.~\ref{tab:noise_values}, we build the sensitivity curves for two cases: Gaia (single-star astrometry), and R-Gaia (relative astrometry), assuming $N_{\rm stars}=10^8$ and $N_{\rm stars}=10^9$ stars (or pairs of stars in the relative case) compressed into $N_{\rm eff}=100$ pixels and $T_{\rm obs}=10$ years of observation. For the relative case, we fix the unperturbed relative angle between stars to $106.5^\circ$, which corresponds to the basi angle between the two FoV of Gaia. The reason for this is that by taking the stars forming each pair within the same FoV, the angular separations $\psi^k_{12}$ will always be $\simeq 10^{-2}$ rad, and as discussed in Sec.~\ref{sec:response_to_sgwb} and Appendix~\ref{app:differential_shift}, the small angle limit is not sensitive and hence irrelevant. The SNR trend as a function of the angle $\psi_{12}$, assumed to be qeual for all the pairs, is shown in Fig. ~\ref{fig:SNR_psi}). 

The result for the sensitivity curves is shown in Fig.~\ref{fig:sensitivity curves}. In the left panel, we see that the relative case performs slightly better than the single-star case, and the sensitivity strongly scales with the number of stars. However, even in the most optimistic scenario with $10^9$ stars, we see in the left panel that Gaia as a GW detector only seems to perform equally well as current PTAs (with 10 years of observation). The combination PTA+Gaia provides a slight improvement to current PTA sensitivity, especially at higher frequencies.  This is because PTA observes timing residuals that behave as $f^3$ in characteristic strain units (see Eq.~\eqref{eq:pulsar_noise}), strongly deteriorating the sensitivity at higher frequencies~\cite{Moore:2017ity}. Combining PTA and Gaia data still represents a considerable advantage when characterizing high frequency features (like pulsar noise) since these detectors are completely independent. This advantage should be seriously considered in future astrometric surveys, in particular if these observe each star with higher cadence compared to Gaia, as a higher cadence of observation would yield even lower levels of noise at high frequency for the same total integration time \cite{Golat_2022}.

In Fig.~\ref{fig:fisher_ptagaia} and ~\ref{fig:fisher_ptagaia_astro}, we show respectively the Fisher uncertainty forecasts for a SGWB signal following a power-law spectrum with (i) $A_{\rm GW}=10^{-14}$ and $\alpha=0$, and (ii) $A_{\rm GW}=3 \times 10^{-15}$ and $\alpha=-2/3$, only considering the case with $N_{\rm stars}=10^9$. The first set of values corresponds to the properties of the GWB that were recently inferred by PTA collaborations~\cite{NANOGrav:2023gor, EPTA:2023fyk, Reardon_2023, Xu_2023} while the second corresponds to what is expected for an astrophysical population of SMBHBs in circular orbit~\cite{Sesana:2008mz}. In scenario (ii), since PTAs outperform Gaia at low frequencies, the improvement in parameter estimation is marginal. This is in agreement with the results in~\cite{Cruz:2024diu}. Such improvement, however, might be more notable when searching for individual SMBHBs or anisotropies in the GWB~\cite{geyer2025influencecontinuousplanegravitational}. In fact, we see in Fig.~\ref{fig:fisher_ptagaia} that the case (i) using the inferred values of parameters (producing a flatter spectrum) gives a more optimistic forecast. This is because the signal is more prominent at higher frequencies where Gaia performs better than PTA. Nevertheless, this is true only when considering $N_{\rm stars}=10^9$. For $N_{\rm stars}=10^8$, in both cases, the improvement is practically null. In Tab.~\ref{tab:SNR_table}, we report the values of the SNR computed for different combined datasets and properties of the SGWB for both scenarios (i) and (ii).

When directly comparing the performance of the relative setup with that of standard single-direction astrometry, we find that the relative approach exhibits significantly lower sensitivity when restricted to star pairs with small angular separations observed simultaneously within the same field of view (FoV). However, for star pairs with large angular separations, the relative method achieves comparable — or even slightly improved — precision. This improvement becomes particularly evident when selecting star pairs located in opposite Gaia FoVs, separated by the spacecraft’s opening angle of $\theta = 106.5^\circ$.
Unfortunately, in order to employ a relative method by making use of the equal-time datastreams in the two Gaia telescopes, we need to keep track of star pairs that are present in the two fields of view \emph{at the same time} for multiple transits during the survey. As explained in Fig.6 of~\cite{Gaia:2016zol}, all the objects surveyed will be individually present multiple times over the duration of the mission on one of the two telescopes. However, in order to %have the condition stated above (
observe a pair of objects simultaneously in both telescopes for multiple consecutive transits, the Gaia spin axis would need to align with the same direction at every subsequent transit, in galactic coordinates. However, the Gaia spin axis is precessing along the Gaia-Sun direction (which is changing in time) with a fixed angle of $45^\circ$, hence this condition is almost never met, making it in practice impossible to have a significant amount of subsequent joint transits of the two telescopes over the same two objects.

It is worth noting that the comparison has been carried out considering a number of $N$ stars versus $N$ star pairs. While the number of possible pairs that can be formed from $N$ stars is significantly larger, Eq.~\eqref{eq:rel_from_abs_1} suggests that the number of independent correlators among these pairs is still $N(N-1)/2$. This makes the assumption of having $N$ independent pairs a reasonable approximation. In practice, however, the effective number of usable stars and star pairs depends on several factors, most notably, the varying measurement accuracy across the catalogue and the spatial distribution of sources on the sky. Accurately assessing these effects would require generating mock datasets that incorporate the specific implementation details of the scanning law and the astrometric solution, which is beyond the scope of this work.

\section{Conclusions}
\label{sec:conclusions}
We have performed a detailed comparison between two astrometric techniques: relative astrometry, based on relative angles between pairs of stars, and single-direction astrometry, where the astrometric shift is measured in a fixed direction in the satellite reference frame. Relative astrometry has been considered because, in principle, it could mitigate systematic errors by circumventing challenges related to the astrometric fit.
This latter approach has also been explored in the context of Gaia~\cite{Crosta:2024udx}, but, as we have discussed extensively in Sec.~\ref{sec:astrometric_response_tot} and Appendix~\ref{app:differential_shift}, that analysis employs an incorrect expression for the relative astrometric shift, which
leads to unrealistically optimistic forecasts.

By correcting the expression used by Ref.~\cite{Crosta:2024udx}, we show that the relative astrometry approach is theoretically well-founded, but its practical application to Gaia data does not yield advantages over the single-star case. In the small-angle limit, the sensitivity of the relative approach is suppressed, with the SNR scaling linearly with the angle between the two stars. In the large-angle regime—where relative astrometry yields forecasts comparable to those of single-direction measurements—the practical application of the relative technique is hindered by the complexity of the Gaia scanning law. The latter makes the pairing of stars located in different fields of view impossible, as discussed in detail at the end of Sec.~\ref{sec:results}. 

Our results also show that astrometric GW measurements with Gaia, when combined with PTA experiments, have the potential to slightly improve SGWB bounds at frequencies $\gtrsim 10^{-7}$ Hz, while their contribution (relative to PTAs) is less significant at lower frequencies.

Our work also leads to an important additional result: the derivation of the astrometric shift, presented in Appendix~\ref{app:astrometric_shift}, which in the limit $f \tau \gg 1$ (with $f$ the typical GW frequency and $\tau$ the light travel time from the star) is substantially simpler than the classical treatments found in~\cite{PhysRevD.83.024024, Pyne:1995iy}.

% Acknowledgments
\begin{acknowledgments}
We acknowledge support from the European Union’s H2020 ERC Consolidator Grant ``GRavity from Astrophysical to Microscopic Scales'' (Grant No. GRAMS-815673, to E.B. and M.V.), the European Union’s Horizon ERC Synergy Grant ``Making Sense of the Unexpected in the Gravitational-Wave Sky'' (Grant No. GWSky-101167314, to E.B.), the PRIN 2022 grant ``GUVIRP - Gravity tests in the UltraViolet and InfraRed with Pulsar timing'' (to E.B. and M.V.), and the EU Horizon 2020 Research and Innovation Programme under the Marie Sklodowska-Curie Grant Agreement No. 101007855 (to E.B.).
A.S. and M.F. acknowledge support European Union’s H2020 ERC Advanced Grant ``PINGU'' (Grant No. 101142079) and from the PRIN 2022 grant "GRAPE" (CUP H53D2300091
0006). G.M. acknowledges support from the Imperial College London Schr\"odinger Scholarship scheme and the hospitality of Milan Bicocca, which provided office space during some parts of this project. A.K. acknowledges funding from the FCT project ``Gravitational waves as a new probe of fundamental physics and astrophysics'' grant agreement 2023.07357.CEECIND/CP2830/CT0003.
\end{acknowledgments}

% Appendices
\appendix

\section{Astrometric shift from conservation equations}
\label{app:astrometric_shift}
Here, we derive the expression for the astrometric shift induced by a plane gravitational wave on light ray propagation. Consider a plane GW described by:
\begin{equation}
h_{00} = h_{0i} = 0\,, \qquad h_{ij} = A_{ij}e^{ik_\mu x^\mu}\,,
\label{eq:wave}
\end{equation}
where $k^\mu = k^t(1, p^i)$ and $p^i = k^i/k^t$ is the propagation direction of the wave ($k^\mu k_\mu = 0 \implies \mathbf{p}\cdot \mathbf{p}=1$). Assuming the wave is in the ``transverse-traceless (TT)" gauge
\begin{equation*}
A_i^i = k^j A_{ij} = 0\,,
\end{equation*}
we take two mutually orthogonal unit vectors $\mathbf{a}$ and $\mathbf{b}$ such that $\mathbf{a}, \mathbf{b} \perp \mathbf{p}$.\footnote{These can be explicitly constructed as
\[
\mathbf{a} = \frac{\hat{\mathbf{y}} \times \mathbf{p}}{|\hat{\mathbf{y}} \times \mathbf{p}|}= \frac{1}{\sqrt{p_x^2 + p_z^2}} \begin{pmatrix}
p_z \\
0 \\
-p_x
\end{pmatrix}\,,
\]
\[
\mathbf{b} = \mathbf{p} \times \mathbf{a}=\frac{1}{\sqrt{p_x^2 + p_z^2}} \begin{pmatrix}
-p_y p_x \\
p_x^2 + p_z^2 \\
-p_y p_z
\end{pmatrix}\,,
\]
and can be used to define polarization tensors
\[
A_{ij}=A_{+}e^{+}_{ij}+A_{\times}e^{\times}_{ij}\,,\qquad \mathbf{e}^{+} = \mathbf{a} \otimes \mathbf{a} - \mathbf{b} \otimes \mathbf{b}\,, \qquad
\mathbf{e}^{\times} = \mathbf{a} \otimes \mathbf{b} + \mathbf{b} \otimes \mathbf{a}\,.
\]}
The spacetime metric $g_{\mu\nu} = \eta_{\mu\nu} + h_{\mu\nu}$, with $h$ given by Eq.~\eqref{eq:wave}, admits three Killing vectors:
\begin{align}
\chi_1 &= a^i\partial_i\,, \nonumber \\
\chi_2 &= b^i\partial_i\,, \nonumber \\
\chi_3 &= \frac{\partial}{\partial t} + p^i\partial_i.
\end{align}

One can show that $\chi_{1}$ is a Killing vectors by evaluating
\[
\mathcal{L}_a g_{\mu\nu} = a^k \partial_k h_{\mu\nu} + \mathcal{O}(h^2) = i k_t \delta_{ij}a^ip^j h_{\mu\nu} + \mathcal{O}(h^2) = \mathcal{O}(h^2)\,.
\]
and similarly for $\chi_2$ and $\chi_3$. We then introduce the following spacetime vectors
\begin{equation}
e_{(0)} = \partial_t + \mathcal{O}(h^2)\,, \quad
e_{(i)} = \partial_{i} - \frac{1}{2}h_{i}^k\partial_{k} + \mathcal{O}(h^2)\,,
\label{eq:tetrad}
\end{equation}
where $h_i^k \equiv \delta^{jk}h_{ij}$. It is straightforward to show that they form a tetrad basis for the spacetime $\eta_{\mu\nu}+h_{\mu\nu}$ 
\begin{align}
g_{\mu\nu}e_{(0)}^\mu e_{(0)}^\nu &= g_{tt} = -1\,, \qquad \\
g_{\mu\nu}e_{(0)}^\mu e_{(i)}^\nu &= g_{it} = 0\,, \qquad \\
g_{\mu\nu}e_{(i)}^\mu e_{(j)}^\nu &= g_{ij} - \frac{1}{2}h_{i}^k\delta_{kj} - \frac{1}{2}h_{j}^k\delta_{ki} + \mathcal{O}(h^2)\,, \nonumber\\
&= \delta_{ij} + \mathcal{O}(h^2)\,.
\end{align}

Now, consider a light ray with wave vector
\begin{equation}
\sigma = \nu e_{(0)} + \nu n^i e_{(i)}\,,
\label{eq:photon_wave_vector}
\end{equation}
where
\[
\mathbf{n} =
\begin{pmatrix}
\sin{\theta}\cos{\phi} \\
\sin{\theta}\sin{\phi} \\
\cos{\theta}
\end{pmatrix}\,.
\]
Here, we leave as implicit the dependence of the photon frequency $\nu$ and propagation direction $n^i$ on the spacetime coordinates $(t,x^i)$.

The Killing vectors $\{\chi_i\}_{i=1}^3$ are associated with the following conservation equations along the ray propagation:
\begin{align}
-g_{\mu\nu}\sigma^\mu\chi_3^\nu &= -g_{\mu\nu}\nu (e_{(0)}^\mu + n^ie_{(i)}^\mu)(e_{(0)}^\nu + p^\nu)\,, \nonumber\\
&= \nu - \nu n^i(g_{\mu\nu}p^\mu e_{(i)}^\nu) = \nu(1 - \mathbf{n} \cdot \mathbf{p}) = \text{const}\,, \label{eq:first}
\end{align}
\begin{align}
g_{\mu\nu}\sigma^\mu \chi_1^\nu &= g_{\mu\nu}\nu (e_{(0)}^\mu + n^ie_{(i)}^\mu) a^\nu \,,\nonumber\\
&= \nu n^i(g_{\mu\nu}a^\mu e_{(i)}^\nu) = \nu \left(\mathbf{n} \cdot \mathbf{a} + \frac{1}{2} h_{ij} n^i a^j\right) = \text{const}\,, \label{eq:second}
\end{align}
\begin{align}
g_{\mu\nu}\sigma^\mu \chi_2^\nu &= g_{\mu\nu}\nu (e_{(0)}^\mu + n^ie_{(i)}^\nu)\udot b\,, \nonumber\\
&= \nu n^i(g_{\mu\nu}b^\mu e_{(i)}^\nu) = \nu \left(\mathbf{n}\cdot \mathbf{b} + \frac{1}{2} h_{ij} n^i b^j \right) = \text{const}\,. \label{eq:third}
\end{align}
where in expanding $g_{\mu\nu}a^\mu e_{(i)}^\nu$ and $g_{\mu\nu}b^\mu e_{(i)}^\nu$ we made use of the fact that for a generic space-like vector $u=u^\mu \partial_{\mu}$ 
\begin{equation*}
g_{\mu\nu}u^\mu e_{(i)}^\nu =u^k e^{(j)}_k(g_{\mu\nu}e_{(j)}^\mu e_{(i)}^\nu)=u^k e^{(j)}_k\delta_{ij}\,,
\end{equation*}
and 
\begin{equation*}
e^{(j)}_k=\left(u^j+\frac{1}{2}h_{k}^{j}u^k\right)\,,
\end{equation*}
is the matrix that connects tetrad and spacetime components. We are ultimately interested in the temporal variation of the direction of arrival of the light ray $\mathbf{n}$ when a GW passes between a star and the Earth.

We thus use the conservation equations~\eqref{eq:first}-\eqref{eq:third} between the star at emission time and the Earth at reception time. In particular, Eq.~\eqref{eq:first} gives
\begin{align}
\nu_E(1-\mathbf{n}_E\cdot \mathbf{p}) = \nu_S(1-\mathbf{n}_S\cdot \mathbf{p}) \; ,
\label{eq:first_cons_eq}
\end{align}
where on the left we have the \textit{Earth terms} (labeled with a $E$) evaluated at the location of Earth at time $t$, while on the right we have the \textit{star terms} (labeled with a $S$) evaluated at the location of the star and at emission time $t_S$. Note that the direction of the GW $p^i$ is the same at Earth and star locations. %as its coordinates don't change in the tetrad frame. 
We can now linearize the difference between Earth and star terms in the perturbation as
\begin{equation*}
\nu_E = \nu_S+\Delta\nu+\mathcal{O}(h^2)\,, \quad \mathbf{n}_E = \mathbf{n}_S+\Delta \mathbf{n}+\mathcal{O}(h^2) \; ,
\end{equation*}
and from Eq.~\eqref{eq:first_cons_eq} we have
\begin{equation}
 \frac{\Delta \nu}{\nu}  = \frac{\Delta \mathbf{n} \cdot \mathbf{p}}{1 - \mathbf{n} \cdot \mathbf{p}}+ \mathcal{O}(h^2)\,.
\label{eq:first_cons}
\end{equation}
Note that we have omitted the $E/S$ subscript on $\nu$ and $\bf n$ in this expression, since it does not make any difference to use one or the other at first order.

Next, we consider the squared sum of Eqs.~\eqref{eq:second} and \eqref{eq:third}:
\begin{align}
(g_{\mu\nu}\sigma^\mu \chi_2^\nu)^2 + (g_{\mu\nu}\sigma^\mu \chi_2^\nu)^2 = \nu^2 \left( 1 - (\mathbf{n} \cdot \mathbf{p})^2 + h_{ij} n^i n^j \right)\,, \nonumber
\end{align}
where in the last line we have used the identity $a^{j} (\mathbf{n} \cdot \mathbf{a}) + b^{j} (\mathbf{n} \cdot \mathbf{b}) + p^j (\mathbf{n} \cdot \mathbf{p}) = n^j$
and the transverseness condition $h_{ij} p^j = 0$. Starting from 
\begin{align}
\nu_E^2 (1 - (\mathbf{n}_E \cdot \mathbf{p})^2 &+ h^E_{ij} n^i n^j) = \nonumber \\
&\nu_S^2 (1 - (\mathbf{n}_S \cdot \mathbf{p})^2 + h^S_{ij} n^i n^j)\,,
\end{align}
and linearizing around $\nu_E, n_E$ we get
\begin{align}
\frac{\Delta \nu}{\nu} = -  \frac{1}{2} \frac{(h_{ij}^E - h_{ij}^S) n^i n^j}{(1 - \mathbf{n}\cdot \mathbf{p})}\ + \mathcal{O}(h^2)\,, 
\label{eq:delta_nu}
\end{align}
where we have used Eq.~\eqref{eq:first_cons} in the last line. 
Eq.~\eqref{eq:delta_nu} is a well-known result in pulsar timing. Then we can write for the direction $\mathbf{n}$
\begin{align*}
\Delta n^i &= \Delta (\mathbf{n} \cdot \mathbf{a}) a^i + \Delta (\mathbf{n} \cdot \mathbf{b}) b^i + \Delta (\mathbf{n} \cdot \mathbf{p}) p^i\,, 
\end{align*}
and using~\eqref{eq:delta_nu} and after some algebra we obtain
\begin{align}
\Delta n^i &= \frac{n^i  +p^i}{2 (1 + \mathbf{n} \cdot \mathbf{p})} (h_{jk}^E - h_{jk}^S) n^j n^k \nonumber \\
&- \frac{1}{2} \big( {h^{i}_{j}}^E - {h^{i}_{j}}^S \big) n^j + \mathcal{O}(h^2)\,.
\label{eq:astrometric_shift_app}
\end{align}
In practical applications, it is possible to neglect the star term from all of the above equations. This is because it typically oscillates very quickly and averages to zero when computing the correlation function of a pair of independent stars. More precisely, let us imagine we have to compute the temporal average of an Earth term, say $f_E(t)$ (where $f$ can be either $\Delta \nu$, $\Delta n^i$ or $h_{ij}$) multiplied with a star term $f_S(t-\tau)$. We have introduced the time $\tau$ that the light ray takes to travel from the star to the Earth. Using the Fourier transform $\tilde f$ of $f$, this reads
\begin{equation}
    \left\langle f_E(t) f_S(t-\tau) \right\rangle = \frac{2 \pi}{T_\mathrm{obs}} \int \mathrm{d}\omega \; e^{i \omega \tau} \tilde f_E(\omega) \tilde f_S(-\omega)\,,
\end{equation}
where $T_\mathrm{obs}$ is the time of observation. Using that $\omega \tau \gg 1$ for the typical frequencies $\omega$ probed in our setting and typical star distances $\tau$, this integral is quickly oscillating and averages to zero. The same is true for the average of two independent star terms $f_{S,1}(t-\tau_1)$ and $f_{S,2}(t-\tau_2)$. To conclude, we will neglect all star terms in the main text.

We have checked that our result in Eq.~\eqref{eq:astrometric_shift_app} reproduces Eq.~(58) in~\cite{PhysRevD.83.024024} once we neglect the star term and replace the direction of propagation of the photon with the direction of the source by $\mathbf{n}\to -\mathbf{n}$ (i.e. $\mathbf{n}_{E/S}\to -\mathbf{n}_{E/S}$ and $\Delta \mathbf{n}\to -\Delta \mathbf{n}$). 
On the other hand, there naively seems to be a discrepancy in the star term between our Eq.~\eqref{eq:astrometric_shift_app} and Eq.~(58) of~\cite{PhysRevD.83.024024}. This is due to different choices for the definition of the astrometric deflection $\Delta \bf n$. While in this Appendix we define $\Delta \bf n$ as the difference between Earth and star terms, in~\cite{PhysRevD.83.024024} $\delta \bf n$ is defined as a purely first-order quantity, i.e. the part of $\bf n$ that oscillates in time at any spacetime point. In order to change between one definition and the other, we have to pay attention to the fact that $\bf n_S$ actually contains a first-order piece that oscillates in time. This is because the direction of the photon at emission time cannot be arbitrary if the photon has to reach Earth following geodesics of the perturbed metric, see~\cite{PhysRevD.83.024024} for more details. Since we will ultimately neglect all star terms in the main text, this technical point does not have any impact on our results, and both definitions are equivalent. This is why we used the notation $\delta \bf n$ for the astrometric deflection in Eq.~\eqref{eq:astrometric_shift}.

\section{Amending previous results for the relative astrometric response}
\label{app:differential_shift}
In Sec.~\ref{sec:astrometric_response}, we derived the relative astrometric response by computing the variation $\delta \cos{\psi_{12}}$ of the angle between a pair of stars identified by the directions $\mathbf{n}_1$ and $\mathbf{n}_2$ on the sky, up to first order in the GW perturbation $h$. We also noted that when employing the relation $\delta\cos{\psi_{12}} \simeq -\delta \psi_{12} \sin{\psi_{12}}$, valid for small unperturbed angles $\psi_{12}$, a direct comparison of Eq.~\eqref{eq:delta_cos_expl} with Eq.~(12) of Ref.~\cite{Crosta:2024udx} reveals their expression lacks the term $- h_{ij} n_{1}^{i} n_{2}^{j}$ on the right-hand side. In the following, we discuss the origin of this discrepancy and explain why this term is necessary to recover the correct limit when $\psi_{12} \sim 0$ (coincident stars limit).

In Ref.~\cite{Crosta:2024udx}, the $\delta \ell^i$ (Eq. (5) there) refer to the variation of the spatial components $\ell^i$ of the photon wave vector in the spacetime basis. In contrast, the standard expression for the astrometric deflection (our Eq.~\eqref{eq:astrometric_shift}) refers to the tetrad components $n^i$, as explained in the main text. The $\delta \ell^i$ are connected to the $\delta n^i$ by the transformation
\begin{equation}
\delta \ell^i = \delta n^i-\frac{1}{2}h_{ij}\ell_{[0]}^j + \mathcal{O}(h^2)\,,
\label{eq:correct_shift}
\end{equation}
where, for improved clarity, we added the $[0]$ subscript to explicitly indicate zero-order quantities in $h$. Since the unperturbed components, being zero-order quantities in $h$, coincide (i.e. $\ell_{[0]}^i=n_{[0]}^i$), one can write
\begin{equation}
\delta \ell^i = \left(\frac{\ell_{[0]}^i +p^i}{2 (1 + \ell_{[0]} \cdot p)} \ell_{[0]}^j \ell_{[0]}^k - \delta^{ik}\ell_{[0]}^j\right) h_{jk}\,.
\label{eq:correct_shift_2}
\end{equation}
In Ref.~\cite{Crosta:2024udx}, the authors replace their Eq.~(11) (equivalent in form to our Eq.~\eqref{eq:astrometric_shift}), in their Eq.~(5), instead of (the correct) Eq.~\eqref{eq:correct_shift_2}. 
Surprisingly, the derivation in the Appendix of~\cite{Crosta:2024udx}, for the variation of the spacetime components $\ell^i$ is correct, in the sense that substituting Eq.~(35) in Eq.~(33) actually gives our Eq.~\eqref{eq:correct_shift_2}. 
However, subsequently, the authors omit a factor of 2 and obtain their Eq.~(12).  

The expression in Eq.~\eqref{eq:delta_cos_expl} for the change in the cosine of the relative angle between the stars in the pair behaves correctly in the limit where the angular separation is small, i.e. $\psi_{12[0]} \sim 0$. If the unperturbed directions of the two stars coincide, the photons from both travel along the same path toward the observer. As a result, any perturbation affects both trajectories identically, and the stars will still appear with zero angular separation,  i.e. $\delta\psi_{12} = 0$. Expanding Eq.~\eqref{eq:delta_cos_expl} around  $\psi_{12[0]} = 0$ (cf. Eq.~\eqref{eq:small_angle}) one finds that $\delta \cos{\psi_{12}}$ is zero up to second order in $\psi_{12[0]}$, which means that $\delta \psi_{12}\sim \delta \cos{\psi_{12}}/\psi_{12[0]}$ goes to zero as well. In contrast, Eq. (12) of~\cite{Crosta:2024udx} is divergent for arbitrarily small angles, resulting in the prediction of an unrealistically high amplitude of the effect, when very small angles (comparable to the resolution power of the telescope) are considered.

\section{Transmission function}
\label{app:transmission_function}

In PTA and Gaia experiments, for observed data $d_0(t)$ in the time domain, we will observe a deterministic trend $s(t)$ that comes from the physical evolution of the object or other processes affecting the observations. Typically, for pulsars, the deceleration of their spin rate with time and, for stars, their proper motion on the celestial sphere. Since most effects are slow with respect to the considered time of observation $T_{\rm obs}$, they can be approximated as a 2nd order polynomial in time as $s(t) = a_0 + a_1t + a_2 t^2 +\mathcal{O}(3)$. In the following, we treat time and frequency as dimensionless quantities with $t \equiv t / T_{\rm obs}$ and $f \equiv f T_{\rm obs}$. We can define an orthogonal basis that spans the space of quadratic polynomials
\begin{equation}
    \Psi=\frac{1}{\sqrt{N}}\ [1,\ \ (2 \ \vec{t})\sqrt{3},\ \ (6\ \vec{t}^2 - \frac{1}{2})\sqrt{5}]\,,
\end{equation}
with $\vec{t}$ the vector of evenly spaced time of observations for $N$ observations spanning between $-1/2$ and $1/2$. The factors $\sqrt{N}$, $\sqrt{3}$ and $\sqrt{5}$ are given by the normalization of the basis. In reality, this basis is only approximately orthonormal because we assume equivalence between integrals and discrete sums for its normalization. Still, the errors are small and can be neglected when $N \gg 1$.

The post-fit residuals are given by
\begin{equation}
    d(\vec{t}) = d_0 (\vec{t}) - \Psi \vec{\alpha}\,,
\end{equation}
where $\vec{\alpha}$ is the vector of best fit coefficients that are given by a least square fit, or using the orthogonality of the basis $\Psi$, directly by the projection of the data $d_0(\vec{t})$ on $\Psi$ as $\vec{\alpha} = \Psi ^\top d_0(\vec{t})$ and
\begin{equation}
    d(\vec{t}) = \left [I_N - \Psi \Psi^\top \right] d_0(\vec{t})\,.
\end{equation}
with $I_N$ the identity matrix of rank $N$. The post-fit residuals are obtained through the application of the projection operator $I_N - \Psi \Psi^\top$ that acts as a filter. Note that $I_N - \Psi \Psi^\top$ is idempotent, meaning that $[I_N - \Psi \Psi^\top][I_N - \Psi \Psi^\top]^\top = I_N - \Psi \Psi^\top$.

Assuming that the data contains Gaussian and stationary noise, the covariance matrix for $d(\vec{t})$ is given by
\begin{equation}
\begin{aligned}
    \langle dd^\top\rangle & = [I_N - \Psi \Psi^\top ] \langle d_0 d_0 ^\top \rangle [I_N - \Psi \Psi^\top]^\top\,,\\
    & \approx \langle d_0 d_0 ^\top \rangle [I_N - \Psi \Psi ^\top]\,,\\
    & \approx \int df S_n(f) \mathfrak{T}(f) e^{i2\pi (t - t')}\,,
\end{aligned}
\end{equation}
where in the second line we have assumed that $\langle d_0 d_0^\top \rangle$ and $[I_N - \Psi \Psi^\top]$ commute, which is true if $d_0$ only contains uncorrelated white noise and $\langle d_0 d_0^\top \rangle$ is diagonal. The presence of time-correlated noise in the data introduces degeneracies with the timing model that do not allow this commutation (for details, see~\cite{Hazboun_2019}). Still, for the sake of forecasting uncertainties, this approximation is reasonable. Finally, in the last line, we used the Wiener-Khinchin theorem to relate the covariance to the power spectral density. The transmission function $\mathfrak{T}(f)$ is defined as the Fourier transform of $[I_N - \Psi \Psi^\top]$
\begin{equation}
\begin{aligned}
    \mathfrak{T}(f) & = \int d(t - t') [I_N - \Psi \Psi^\top] e^{-i2\pi f (t - t')}\,,\\
    & \approx 1 - \sum^3_{a=1}\frac{1}{N} \Psi_a (t_i) e^{-i 2\pi f t_i} (\Psi_a (t_j) e^{i2 \pi f t_j})^\top]\,, \\
    & \approx 1 - N \sum^3_{a=1} |\tilde{\Psi}_a(f)|^2\,. % \tilde{\Psi}_a^\dagger(f)
\end{aligned}
\label{eq:transmission_fourier}
\end{equation}
Assuming that the spectral density is concentrated on the diagonal $f=f'$ (see~\cite{Hazboun_2019}) and using the equivalence between the continuous and discrete Fourier transforms $\tilde{\Psi}_a(f) \approx \frac{1}{N} \Psi_{a}(\vec{t}) e^{-i2\pi f \vec{t}}$ with $\tilde{\Psi}_a(f)$ the Fourier transform of $\Psi_a(t)$.

The Fourier transform of the $n$-th power of $t$ is given by
\begin{equation}
    \int_{-1/2} ^{1/2} t^n e^{-i2\pi f t} dt = \frac{1}{(-i 2 \pi)^n} \frac{d^n}{df^n} \left[\frac{\sin (\pi f)}{\pi f} \right]\,,
\end{equation}
yielding
\begin{equation}
\begin{aligned}
    \tilde{\Psi}(f) = & \frac{1}{\sqrt{N}} \bigg[\frac{\sin(\pi f)}{\pi f}\,, \\
    & \frac{\sqrt{3}}{-i2\pi} \left(- 2\frac{\sin (\pi f)}{\pi f^2} + 2\frac{\cos(\pi f)}{f}\right)\,,\\
    & \frac{\sqrt{5}}{-4\pi^2} \left( 12 \frac{\sin (\pi f)}{\pi f^3} - 12 \frac{\cos (\pi f)}{f^2} - 4\pi \frac{\sin (\pi f)}{f}\right)
    \bigg]\,.
\end{aligned}
\label{eq:psi_fourier}
\end{equation}

The transmission function is directly given by the squared modulus of $\tilde{\Psi}(f)$ according to Eq.~\eqref{eq:transmission_fourier}. At low frequencies, the behavior of the function is dominated by $f^6$ terms, as highlighted in~\cite{Hazboun_2019, forecasting_pta_gwb} and the operator $I_N - \Psi \Psi^\top$ is acting as a high-pass filter on the data.

Note that this does not reproduce higher frequency features of the transmission function, like the one year peak that is often visible in PTA sensitivity curves or upper limit plots \cite{ipta_dr2_cgw, Hazboun_2019, ng_11yr_gwb}. This peak comes from the pulsar position correction fit in the timing model, that varies periodically with the orbit of the Earth with a one year period. Accounting for this effect leads to a suppression of sensitivity at frequency $f_{\rm yr}=1/(1 \rm yr)$. In this work, we do not implement it since we focus on the low frequency range where it has a negligible impact because $T_{\rm obs}$ is sufficiently greater than one year. It could however easily be implemented by hand with a component $\propto \sin(2\pi f_{\rm yr}t)$ in the basis $\Psi$, assuming that it is orthogonal to other low frequency components since $T_{\rm obs} \gg 1 \rm yr$. This produces a component $(1 - \epsilon) \textrm{sinc} [\pi (f - f_{\rm yr})]/\sqrt{N}$ in frequency domain for $\tilde{\Psi}(f)$. The $\epsilon \approx 10^{-3}$ is introduced to avoid numerical divergences.
% References
\bibliographystyle{apsrev4-1}
\bibliography{refs}

\end{document}